\documentclass[11pt,preprint]{aastex}
\usepackage{graphics}
\usepackage{epsfig}
\usepackage{longtable}
\def\gsim{\lower 2pt \hbox{$\, \buildrel {\scriptstyle >}\over
{\scriptstyle \sim}\,$}}
\def\lsim{\lower 2pt \hbox{$\, \buildrel {\scriptstyle <}\over
{\scriptstyle \sim}\,$}}

\shortauthors{}
\shorttitle{}

\begin{document}

\title{X-ray emission from the Sombrero galaxy: discrete sources}
\author{Zhiyuan Li\altaffilmark{1}, Lee R.~Spitler\altaffilmark{2}, 
  Christine Jones\altaffilmark{1}, William R.~Forman\altaffilmark{1}, \\ 
   Ralph P.~Kraft\altaffilmark{1}, Rosanne Di Stefano\altaffilmark{1},
  Shikui Tang\altaffilmark{3}, Q.~Daniel
  Wang\altaffilmark{3}, \\
  Marat Gilfanov\altaffilmark{4}, Mikhail Revnivtsev\altaffilmark{5,6}} 
\altaffiltext{1}{Harvard-Smithsonian Center for Astrophysics, 60
  Garden Street, Cambridge, MA 02138, USA; zyli@cfa.harvard.edu}
\altaffiltext{2}{Centre for Astrophysics and Supercomputing, Swinburne University, Hawthorn, VIC 3122, Australia}
\altaffiltext{3}{Department of Astronomy, University of Massachusetts,
  710 North Pleasant Street, Amherst, MA 01003, USA}
\altaffiltext{4}{Max-Planck-Institut f${\rm \ddot{u}}$r Astrophysik,
  Karl-Schwarzschild-Str 1, 85741 Garching bei M${\rm \ddot{u}}$nchen, Germany}
\altaffiltext{5}{Excellence Cluster Universe, Technische Universit\"at M\"unchen, Boltzmannstr.2, 85748 Garching, Germany}
\altaffiltext{6}{Space Research Institute, Russian Academy of
  Sciences, Profsoyuznaya 84/32, 117997 Moscow, Russia}

\begin{abstract}
We present a study of discrete X-ray sources in and around the
bulge-dominated, massive Sa galaxy, Sombrero (M104), based on new and archival {\sl Chandra}
observations with a total exposure of $\sim$200 ks. 
With a detection limit of $L_{\rm X} \approx
10^{37}{\rm~ergs~s^{-1}}$ and a field of view covering a
galactocentric radius of $\sim$30 kpc (11\farcm5), 383 sources are
detected. 
Cross-correlation with Spitler et al.'s catalogue of Sombrero globular clusters (GCs) identified from {\sl HST}/ACS observations reveals 
41 X-rays sources in GCs, presumably low-mass
X-ray binaries (LMXBs). Metal-rich GCs are found to have a higher
probability of hosting these LMXBs, a trend similar to that found in
elliptical galaxies. On the other hand, the four most luminous GC LMXBs, with apparently
super-Eddington luminosities for an accreting neutron star, 
are found in metal-poor GCs. 
We quantify the differential luminosity functions (LFs) for both
the detected GC and field LMXBs, whose power-low indices ($\sim$1.1
for the GC-LF and $\sim$1.6 for field-LF) are consistent with
previous studies for elliptical galaxies. 
With precise sky positions of the GCs without a detected X-ray source, we
further quantify, through a fluctuation analysis, 
the GC LF at fainter luminosities down to 
$10^{35}{\rm~ergs~s^{-1}}$. The derived index rules out a faint-end slope
flatter than 1.1 at a 2 $\sigma$ significance, contrary to recent findings in several elliptical galaxies and the
bulge of M31.
On the other hand, the 2-6 keV unresolved emission places a tight constraint 
on the field LF, implying a flattened index of
$\sim$1.0 below $10^{37}{\rm~ergs~s^{-1}}$.
We also detect 
101 sources in the halo of Sombrero. The presence of these sources cannot 
be interpreted as galactic LMXBs
whose spatial distribution empirically follows the starlight. Their number is also
higher than the expected number of cosmic AGNs ($52\pm11$ [1 $\sigma$]) whose surface density 
is constrained by deep X-ray surveys. We suggest that either the cosmic X-ray background is unusually high in the direction of Sombrero, or a distinct population of X-ray sources is
present in the halo of Sombrero.
\end{abstract}

\section{Introduction}
The {\sl Chandra X-ray Observatory}, with its superb angular
resolution and sensitivity, has established the ubiquity
of LMXBs in nearby early-type galaxies (e.g., Sarazin et al.~2000;
Kraft et al.~2001), in particular confirming the hypothesis that such a
population accounts for the bulk of X-ray emission in X-ray-faint 
elliptical/S0 galaxies (i.e., galaxies with a low
$L_X/L_B$ ratio). The total number and
cumulative luminosity of LMXBs are further shown to be  
good indicators of the host galaxy's stellar mass (Gilfanov
2004), with only a weak dependence on morphological type. 
This dependence is at least partly related to the specific
frequency of GCs in which LMXBs are efficiently formed through dynamical
processes (Clark 1975). 

{\sl Chandra} observations of more than a dozen
nearby elliptical/S0 galaxies have provided valuable information on
their LMXB populations (Kim et al.~2006; Kundu, Maccarone \& Zepf~2007; Sivakoff
et al.~2008; Brassington et al.~2008, 2009;
Posson-Brown et al.~2009; Voss et al.~2009; Kim et al.~2009, among many
recent studies). However, a comparative view for bulges of spiral galaxies is still limited.
Our present knowledge comes primarily
from LMXBs detected in M31 (Kong et al.~2002; Voss \& Gilfanov 2007), M81 (Tennant et
al.~2001), M104 (Di Stefano et al.~2003) and our own Galaxy.
Since observations of elliptical/S0 galaxies rarely reach a
source detection limit below
$10^{37}{\rm~ergs~s^{-1}}$, 
LMXBs with luminosities as low as 
$10^{35}$-$10^{36}{\rm~ergs~s^{-1}}$ are chiefly detected in the bulges
of M31 and our Galaxy. However, the number of LMXBs in these two
bulges is limited by their moderate stellar mass
and their relatively low specific GC frequency. Sensitive X-ray studies of 
nearby bulges can improve our knowledge of the LMXB population  
in this important galactic component.
 
The Sombrero galaxy (M104; NGC 4594), 
a bulge-dominated, edge-on Sa galaxy at a distance of 9.0$\pm$0.1 Mpc
(Spitler et al.~2006), 
has a total stellar mass
comparable to that of the most massive elliptical galaxies 
in the Virgo
cluster. The galaxy also harbors a sizable population of GCs,
with an estimated number of $\sim$2000 (Rhode \& Zepf 2004).
Therefore a large number of
LMXBs is expected in Sombrero, making it an ideal target for 
studying both sources residing in the GCs and in the field. 
Indeed, in a shallow {\sl Chandra} observation of Sombrero,
Di Stefano et al.~(2003) found more than 100 discrete sources. 
Two deep {\sl Chandra} observations were recently taken, effectively
adding Sombrero to a short list of early-type galaxies for which a
source detection limit below $10^{37}{\rm ergs~s^{-1}}$ is achieved.
In this work we study the stellar populations, mostly LMXBs, 
based on these observations. In a forthcoming paper we will present
a study of the diffuse X-ray emission.

In \S~\ref{sec:data} we briefly describe the data preparation.
We present our analysis in \S~\ref{sec:anal} and
discuss the results in \S~\ref{sec:dis}. A summary is given in 
\S~\ref{sec:sum}. We quote 1 $\sigma$ errors throughout this work.

\section{Data preparation} {\label{sec:data}}
\subsection{X-ray and optical data}
\begin{figure*}[!htb]
\vskip-2.5cm
\centerline{
    \epsfig{figure=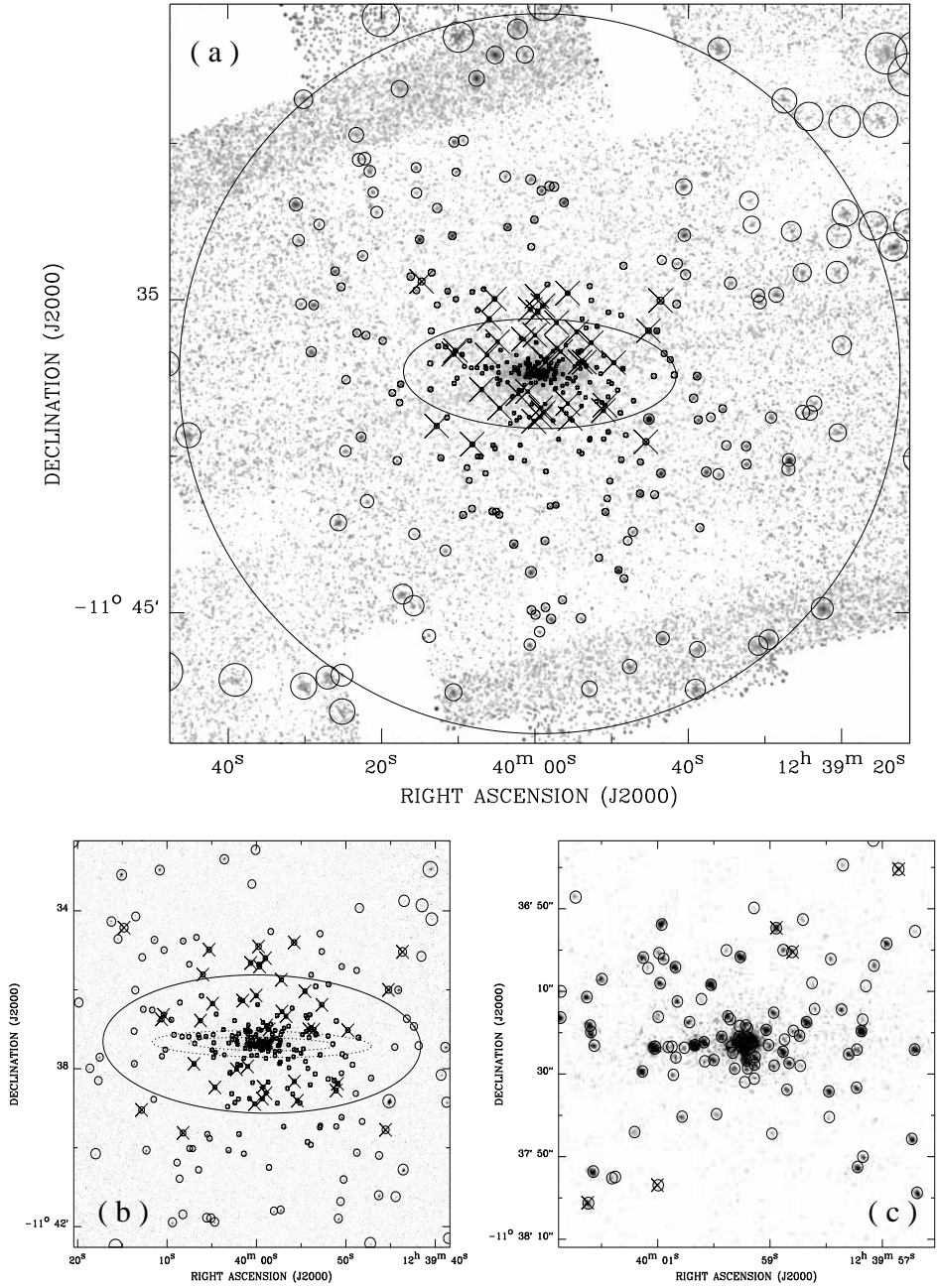,width=0.8\textwidth,angle=0}
    }
  \caption{Smoothed 0.4-6 keV intensity images of M104 in (a) the 23$^\prime$ by
    23$^\prime$ region. Detected X-ray sources are marked
with circles of the 90\% EER. GC sources 
are further marked with crosses. 
The ellipse (8\farcm7 by 3\farcm5) represents the $D_{25}$ isophote
of the galaxy; the large circle encloses sources of interest; (b) 
the $10^{\prime}$ by $10^{\prime}$ region. The two dashed
ellipses outline the dust lane; (c) 
the central 1\farcm5 by 1\farcm5 region.
}
 \label{fig:sou_plot}
\end{figure*}

Existing {\sl Chandra} studies of X-ray sources in Sombrero (Di Stefano et
al. 2003; Kundu et al.~2007; Li, Wang \& Hameed 2007) are based on a 19-ks
ACIS-S observation taken on May 31, 2001 (Obsid.~1586; PI:
S. Murray). On April 29 and December 2, 2008, we obtained two new {\sl Chandra} ACIS-I
observations (Obsid.~9532 and 9533; PI: C.~Jones), with exposures of 
85 and 90 ks, respectively. In this work we utilize data from all three observations. 
We reprocessed the data using CIAO v.4.1 and the corresponding calibration 
files, following the {\sl Chandra} ACIS data analysis guide. 
No time intervals of high particle background were found.
We calibrated the relative astrometry among the three observations
using the CIAO tool {\sl reproject\_aspect}, by matching centroids
of discrete sources commonly detected in all three
observations. The resultant relative astrometry is better than 0\farcs1. 
For each observation, we produced count and exposure
maps in the 0.4-0.7 (S1), 0.7-1 (S2), 1-2 (H1) and 2-6 (H2) keV bands. 
An absorbed power-law spectrum,
with a photon-index of 1.7 and an absorption column density
$N_{\rm H}=10^{21}{\rm~cm^{-2}}$ (a value somewhat higher than the
Galactic foreground column density toward Sombrero,
$3.7{\times}10^{20}{\rm~cm^{-2}}$ [Dickey \& Lockman 1990], but allowing
for some internal absorption), was adopted to calculate spectral weights when
producing the exposure maps. The energy-dependent difference of 
effective area between
the ACIS-S3 CCD and the ACIS-I CCDs was taken into account, assuming
the above incident spectrum, so that
the quoted count rates throughout this work refer to ACIS-I. 
The count and exposure maps of individual observations were projected to a common
tangential point, here the optical center of Sombrero, to produce summed
images of the combined field of view (FoV; Fig.~\ref{fig:sou_plot}a).  
The total effective exposure, in the 2-6
keV band for example, is $\gtrsim$180 ks within a projected
galactocentric
radius $R \approx 4^\prime$, where the FoV is common to the
three observations, and gradually drops below 80 ks at
$R \gtrsim 8^\prime$. 

Hubble Space Telescope ({\sl HST}) Advance Camera for Surveys
(ACS) data in the ``science-ready form''
were used by Spitler et al.~(2006; hereafter S06) to identify GC
candidates in a $\sim$600$^{\prime\prime}\times400^{\prime\prime}$ 6-pointing mosaic centered on Sombrero.
Three optical bands ($BVR$) were obtained through the Hubble Heritage
Project (PI: K.~Noll, PID.~9714). 
S06 used photometric and size-selection
(GCs are partially-resolved on the ACS imaging) 
to construct a catalogue of 659 GC candidates. Owing to the excellent
quality 
of the {\sl HST} images, the
catalogue contains all but the faintest 5\% of the GCs
in this field, with minimal contamination.  
In particular, objects falling on Sombrero's prominent dust lane are excluded. 
Here we use 
improved photometric measurements for the GC candidates (see discussion
in Spitler, Forbes \& Beasley 2008). 

\subsection{X-ray source detection} {\label{subsec:detect}}
Following the procedure detailed in Wang (2004), we detected sources
in the soft 
(S1; 0.4-0.7 keV), medium (M=S2+H1; 0.7-2 keV), hard (H2; 2-6 keV) and full
(F=S1+S2+H1+H2) bands. 
The low- and high-energy cut-offs were chosen to optimize
the signal-to-noise ratio against the instrumental background.
For the full combined FoV, a 2-pixel 
($\sim$1$^{\prime\prime}$) binning was adopted to save computational effort; 
for the inner region ($R\lesssim$4$^\prime$), where source
crowding is expected, the detection was refined using the original
pixel scale ($\sim$0\farcs5/pixel).   
With a local false detection probability $P \leq 10^{-6}$ 
(yielding approximately one false source in the inner region and two false sources
in the full FoV), we detected
a total of 383 sources within $R$=11\farcm5 over the
combined images (Fig.~\ref{fig:sou_plot}). Table \ref{tab:sou}
summarizes the detection results. For each source, background-subtracted, exposure 
map-corrected count rates in individual bands are derived from within the 90\% enclosed
energy radius (EER). We also calculated hardness ratios for each source using the 
method of Bayesian estimation (Park et al.~2006).
In addition, we detected sources in individual observations to study
long-term source variability (see \S~\ref{subsec:var}). Lists of
sources detected in individual observations are summarized in
Tables~{\ref{tab:sou_1586}}, {\ref{tab:sou_9532}} and {\ref{tab:sou_9533}}.

To identify interlopers (e.g., relatively bright foreground stars and
background galaxies), we rely on the USNO-B and
2MASS source catalogs (Monet et al.~2003; Cutri et al.~2003). 
A matching radius of 1$^{\prime\prime}$ is adopted, which yields random
matches of 0.8 X-ray/optical pairs and 0.3
X-ray/near-infrared (NIR) pairs, under the assumption that
interlopers are uniformly distributed within $R$=11\farcm5. 
In light of the intrinsic astrometry offset among the X-ray, optical
and NIR source catalogs, we performed the matching procedure
iteratively, in each run shifting the sky positions (R.A. and Dec.) 
of the optical (or NIR) sources by minimizing the cumulative position 
difference of the matched pairs, until the required shift is less than 0\farcs1. 
A total of 28 X-ray sources were thus identified as interlopers and excluded from
further analysis. We also exclude the nucleus of Sombrero, the
brightest source detected in the FoV (source 194 in Table \ref{tab:sou}). A detailed study of the nucleus
will be given elsewhere.
The remaining 354 sources are the subject of the analysis presented here.

\begin{figure*}[!htb]
\centerline{
    \epsfig{figure=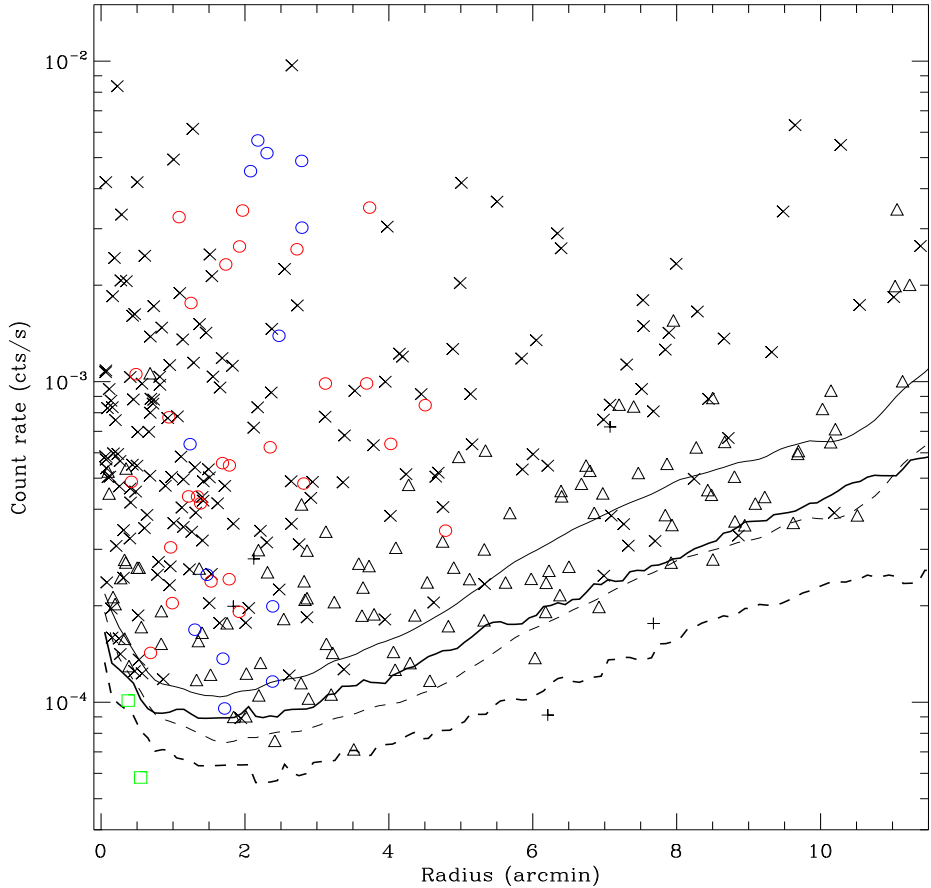,width=0.8\textwidth,angle=0}
  }
\caption{~0.4-6 keV (F-band) count rate vs. galactocentric radius for the detected
  sources. {\sl Red circle}: sources associated with red GCs; 
{\sl blue circle}: sources associated with blue GCs; 
{\sl green square}: sources not detected in the M- and H2-bands;
{\sl triangle}: sources not detected in the S1- and H2-bands;
{\sl plus}: sources not detected in the S1- and M-bands;
{\sl cross}: sources detected in at least three bands.
The solid and dashed curves illustrate the detection threshold in the
F-band and M-band, respectively, with  
  the thin one showing the azimuthal average value and the thick
  one showing the minimum value at each radius.
}
\label{fig:sou_dist}
\end{figure*}

The F-band count rates of individual sources are plotted against
their galactocentric radii in Fig.~\ref{fig:sou_dist},
along with 
the radial variation of the detection threshold that depends on the
local effective exposure, point spread function (PSF) and background.
Across the FoV we achieve a minimum detection threshold of
$8\times10^{-5}{\rm~cts~s^{-1}}$ in the F-band, which corresponds to an
intrinsic 0.5-8 keV luminosity of $\sim$$8\times10^{36}{\rm~ergs~s^{-1}}$.
Here we adopt an F-band count rate-to-flux conversion of
$1.0\times10^{-11}{\rm~ergs~s^{-1}~cm^{-2}}/({\rm cts~s^{-1}})$, or a
F-band count rate-to-luminosity conversion of 
$9.7\times10^{40}{\rm~ergs~s^{-1}}/({\rm cts~s^{-1}})$, for
the assumed absorbed power-law model 
and distance of 9.0 Mpc (1$^\prime$=2.6 kpc) for Sombrero.


\section{Analysis and results} {\label{sec:anal}}
\subsection{X-ray sources associated with globular clusters} {\label{subsec:GC}}
We begin our analysis by cross-correlating the detected X-ray sources with 
the GC catalogue of S06,
with a similar procedure as we described above for the interlopers. A
more conservative matching radius of 0\farcs5 is applied, for which 0.5
random matches are expected. This is estimated by artificially
shifting the positions (R.A. and Dec.) of the X-ray
sources by $\pm5^{\prime\prime}$ and averaging the number of coincident matches. 
41 pairs of GC-X-ray
sources are identified (Fig.~\ref{fig:sou_plot}). Increasing the 
matching radius to 1$^{\prime\prime}$ results in 45 pairs, but
accordingly $\sim$4
are expected to be random matches. 
Given their typical luminosities ($\gtrsim 10^{37}{\rm ergs~s^{-1}}$), these X-ray sources are
most likely LMXBs. For clarity, we refer to them as GC-LMXBs hereafter, which are tabulated in Table~\ref{tab:GCLMXB}.

\begin{figure*}[!htb]
\centerline{
    \epsfig{figure=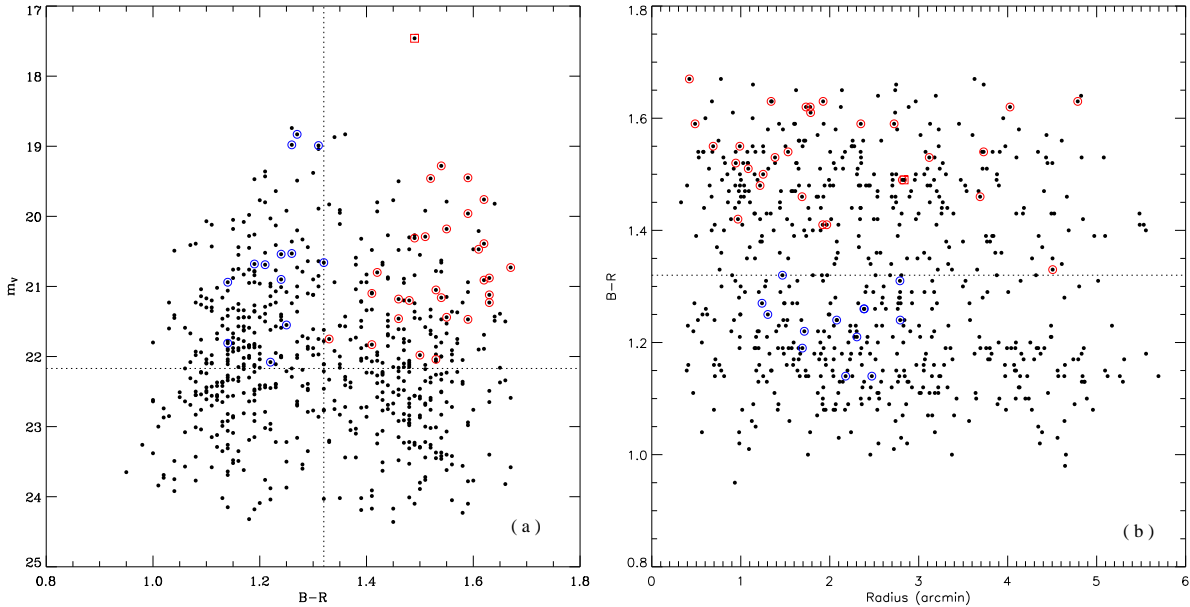,width=\textwidth,angle=0}
   }
\caption{(a) Color-magnitude diagram of all GCs identified in
  S06. Improved photometry is obtained from Spitler, Forbes \& Beasley
  (2008). GCs hosting a detected X-ray source are marked by an open
  circle. The vertical dashed line represents the division of blue ($B-R \leq
1.32$) and red ($B-R>1.3$) GCs, the horizontal line the turnover
magnitude, following S06. Also plotted is the metal-rich UCD (Hau et al.~2009), with
m$_{\rm V}$=17.46, which is found to host an X-ray source and marked
by a red square. (b) GC color vs. galactocentric radius.}
\label{fig:GC}
\end{figure*}

Among the 41 GC-LMXBs, 13 are found in blue (i.e., metal-poor) GCs and 28 in
red (i.e., metal-rich) GCs (Fig.~\ref{fig:GC}).
The entire GC catalogue of S06 harbors a blue-to-red number ratio of 357:302
(54\% are blue). 
Hence the LMXB detection rate is 
3.6$\pm$1.2\%, 9.3$\pm$2.3\% and 6.2$\pm$1.2\% for the blue, red and
total GCs, respectively.
We note that all 41 X-ray sources are identified
with GCs brighter than the V-band turnover magnitude of
$m_{\rm V}$ = 22.17 (Fig.~\ref{fig:GC}a; S06), above which the
blue-to-red GC number ratio is 205:156 (57\% are blue). Considering only GCs brighter
than the turnover magnitude, the LMXB detection rate becomes
5.8$\pm$2.0\%, 17.7$\pm$4.8\% and 11.0$\pm$2.3\% for the blue, red and
total GCs, respectively.
We conclude that in Sombrero metal-rich GCs are more likely to host
a LMXB, a now familiar trend established in 
studies of nearby early-type galaxies (e.g., Kim
et al. 2006; Kundu et al.~2007), although the number ratio between
blue and red GC hosts in Sombrero ($\sim$1:2.2) appears 
higher than the average ratios of 1:3.4 derived by Kundu et al.~(2007) and
1:2.7 by Kim et al.~(2006). 
We caution that the GC-LMXBs are limited in number, and furthermore,
a direct comparison of these numbers might not be warranted
due to the inhomogeneous data and the 
different criteria of selecting GCs adopted by different authors.

We probe a galactocentric radius dependency for the GC-LMXB connection
in Fig.~\ref{fig:GC}b. Both red and blue subpopulations of the S06
GCs are identified out to $R\sim5\farcm7$, with the red GCs more
centrally concentrated than the blue ones (S06), a trend generally
found in elliptical galaxies. Fig.~\ref{fig:GC}b also reveals that
the blue GCs hosting LMXBs are 
located only in the radial range of 1$^\prime$-3$^\prime$. This is 
in contrast with the locations of the red hosts, which
apparently sample the entire radial range.

\begin{figure*}[!htb]
\centerline{
    \epsfig{figure=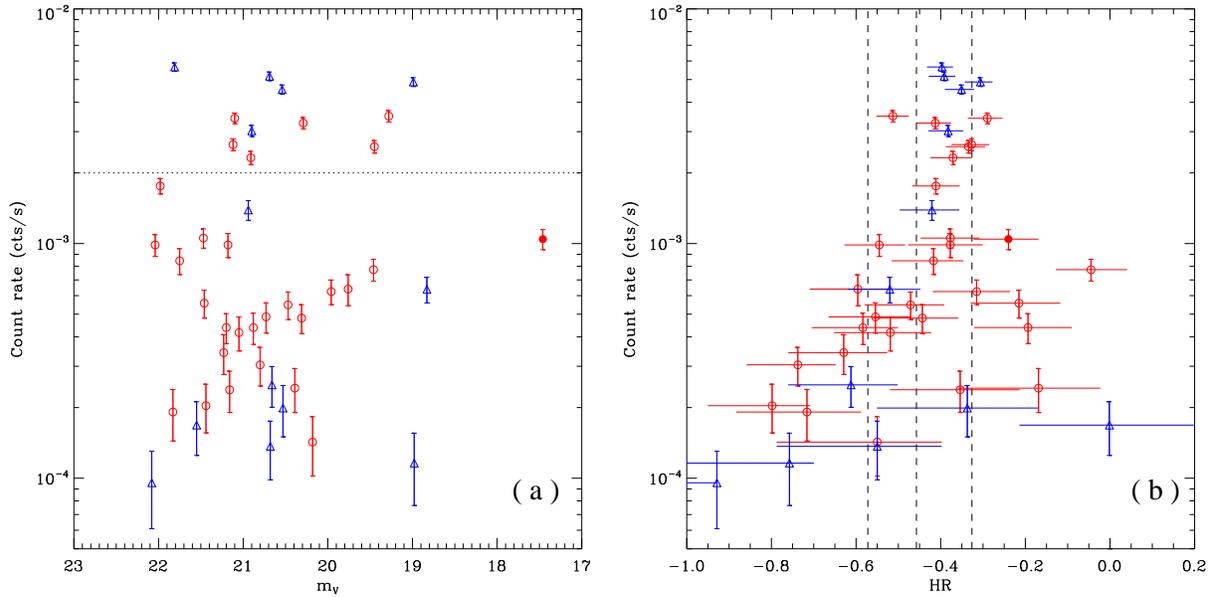,width=\textwidth,angle=0}
   }
\caption{0.4-6 keV count rate vs. (a) V-band magnitude and (b) 
hardness ratio, for the GC-LMXBs. The hardness ratio is defined 
as HR = (H2-M-S1)/(H2+M+S1). Blue and red GCs are plotted with blue triangles and 
red open circles, respectively. The UCD is plotted with a filled circle for comparison.
The horizontal dashed line in (a) represents the Eddington luminosity of an
accreting neutron star with a mass of 1.4 M$_{\odot}$. The vertical
dashed lines in (b), from left to right, correspond to hardness ratios 
for an absorbed power-law spectrum with a fixed $N_{\rm
  H}=10^{21}{\rm~cm^{-2}}$ and a photon-index of 2.0, 1.7 and 1.4, respectively.
}
\label{fig:XGC}
\end{figure*}

\begin{figure*}[!htb]
\centerline{
    \epsfig{figure=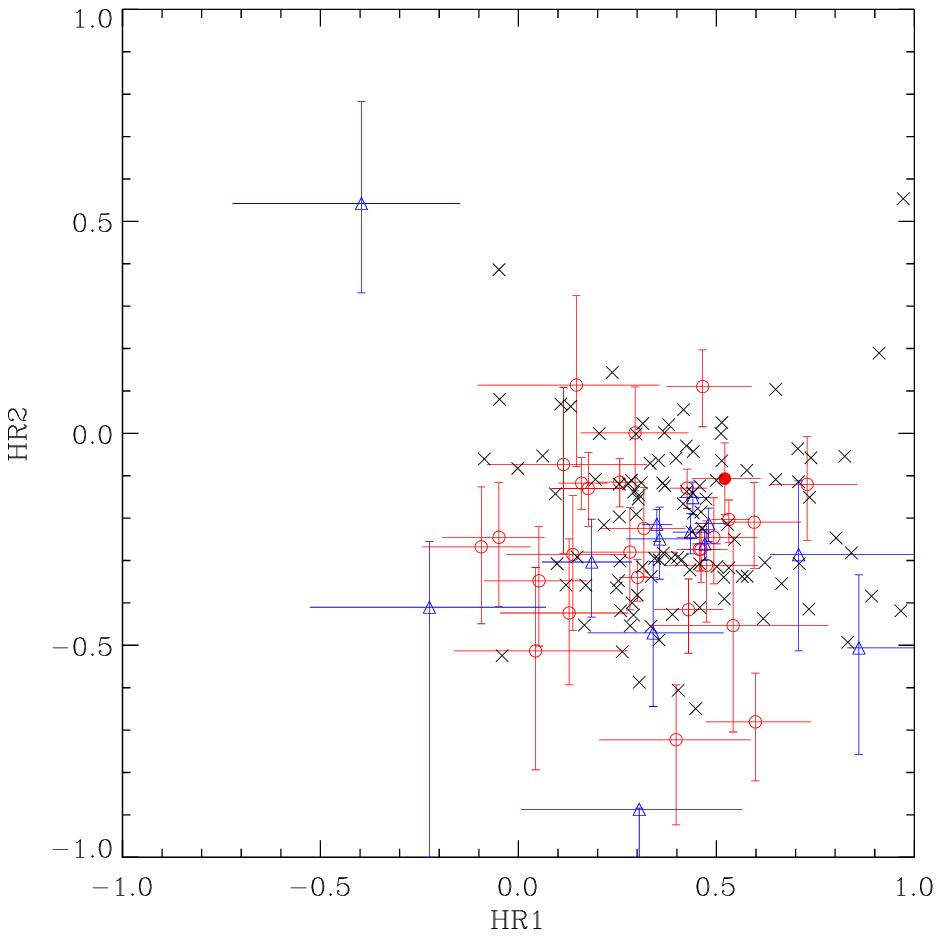,width=0.6\textwidth,angle=0}
   }
\caption{HR1 versus HR2, where HR1 = (H1-S2-S1)/(H1+S2+S1) and HR2 = (H2-H1)/(H2+H1). The blue and red symbols are the same representation of GC-LMXBs as in Fig.~\ref{fig:XGC}a., whereas black crosses represent those field X-ray sources detected within $R=4^\prime$ and simultaneously detected in the F-, M- and H2-bands. For clarity, error bars are not shown for the field sources, which have a similar uncertainty range as those of the GC-LMXBs.}
\label{fig:HR}
\end{figure*}

Fig.~\ref{fig:XGC} shows the F-band count rate and hardness ratio, defined 
as HR = (H2-M-S1)/(H2+M+S1), of
the GC-LMXBs. Eleven sources have apparent luminosities above the
Eddington luminosity for accreting neutron stars
(NSs; $L_{\rm Edd}$$\sim$2$\times10^{38}{\rm~ergs~s^{-1}}$). Among these, 
the four most luminous sources, with luminosities
of $\sim$2-3 $L_{\rm Edd}$, are found in
blue GCs. 
Each of these sources may be the superposition of several accreting NSs or an accreting black
hole (BH), although theoretical considerations of dynamical
processes in GCs predict the ejection of nearly all stellar-mass
BHs on a short timescale of $\sim$$10^8{\rm~yr}$ (Kulkarni, Hut \& McMillan 1993; Sigurdsson \& Hernquist 1993). We note that
the only accreting BH previously found in a GC is located in a blue
GC in NGC~4472 (Maccarone et al.~2007). The GC probably lacks an
intermediate-mass BH to dynamically eject stellar-mass BHs (Zepf et
al.~2008).
 As Fig.~\ref{fig:XGC}a shows,
bright Sombrero GCs (m$_{\rm V}\leq$20) tend to host relatively bright LMXBs,
whereas the LMXBs found in fainter GCs spread over the entire
luminosity range ($\sim$$10^{37}-5\times10^{38}{\rm~ergs~s^{-1}}$). 
In fact, the brightest LMXB is associated with one of the faintest GCs
hosting an X-ray source. 
No significant trend is found in the X-ray flux-hardness ratio distribution of the
GC-LMXBs (Fig.~\ref{fig:XGC}b). 
The brightest sources (those with luminosities 
above $\sim$$10^{38}{\rm~ergs~s^{-1}}$)
show hardness ratios typical of accreting NSs (with photon-indices of 
$\sim$1.4-2). On the
other hand, the fainter sources exhibit a variety of hardness ratios
ranging from very soft ($\sim$-1.0) to very hard ($\sim$0.0).
Analogous to a color-color diagram, Fig.~\ref{fig:HR} plots HR1 versus HR2 for the GC-LMXBs, 
where we define HR1 = (H1-S2-S1)/(H1+S2+S1) and HR2 = (H2-H1)/(H2+H1). Field X-ray sources (i.e., those not associated with GCs) detected within $R=4^\prime$ are also plotted for comparison.
No distinct spectral behavior is evident between the blue and red GC-LMXBs, nor between 
the GC and field sources.

We also note that an ultra-compact dwarf (UCD) associated with
Sombrero was identified in the {\sl HST}/ACS images (Hau et al.~2009), which in many aspects appears as a ``giant version'' of a GC.
We confirm the positional coincidence
between the UCD and an X-ray source (source 275 in
Table \ref{tab:sou}).
The centroids
of the two objects are separated by $0\farcs4\pm0\farcs4$, a value
comparable with the optical
half-light radius ($\sim$0\farcs33) of the UCD. We show this pair of sources
in Fig.~\ref{fig:XGC} and Fig.~\ref{fig:HR} in comparison with the GC-LMXBs. The UCD has
a 0.5-8 keV luminosity of $\sim$$10^{38}{\rm~ergs~s^{-1}}$ and a relatively
soft spectrum (showing HR $\approx$ -0.25) compared to most of the GC-LMXBs.

\subsection{Spatial properties} {\label{subsec:spatial}}
The spatial distribution of detected X-ray sources shows a clear concentration
within the optical extent of the galaxy (Fig.~\ref{fig:sou_plot}b, c). 
The well known presence of a prominent dust lane and its
association with radio continuum emission (Bajaja et al.~1988)
and H$\alpha$ emission (Li et al.~2007) indicate
star-forming activities in the disk. Both the radio continuum and
H$\alpha$ fluxes suggest a rather low star formation rate of 
0.1-0.2 M$_\odot$/yr, which in turn predicts $\leq$4
high-mass X-ray binaries (HMXBs) with luminosities $\geq10^{37}{\rm~ergs~s^{-1}}$
 (Grimm, Gilfanov \& Sunyaev 2003). 
Hence the bulk of sources
associated with the galaxy are presumably LMXBs.
We further note that two sources are only detected in the S-band
(sources 159 and 189 in Table \ref{tab:sou}), suggesting that they are super-soft sources (SSS;
e.g., Di Stefano \& Kong 2004). As SSSs are thought be a different
population than LMXBs, we do not include these two sources
in the following analysis. 

\begin{figure*}[!htb]
\centerline{
    \epsfig{figure=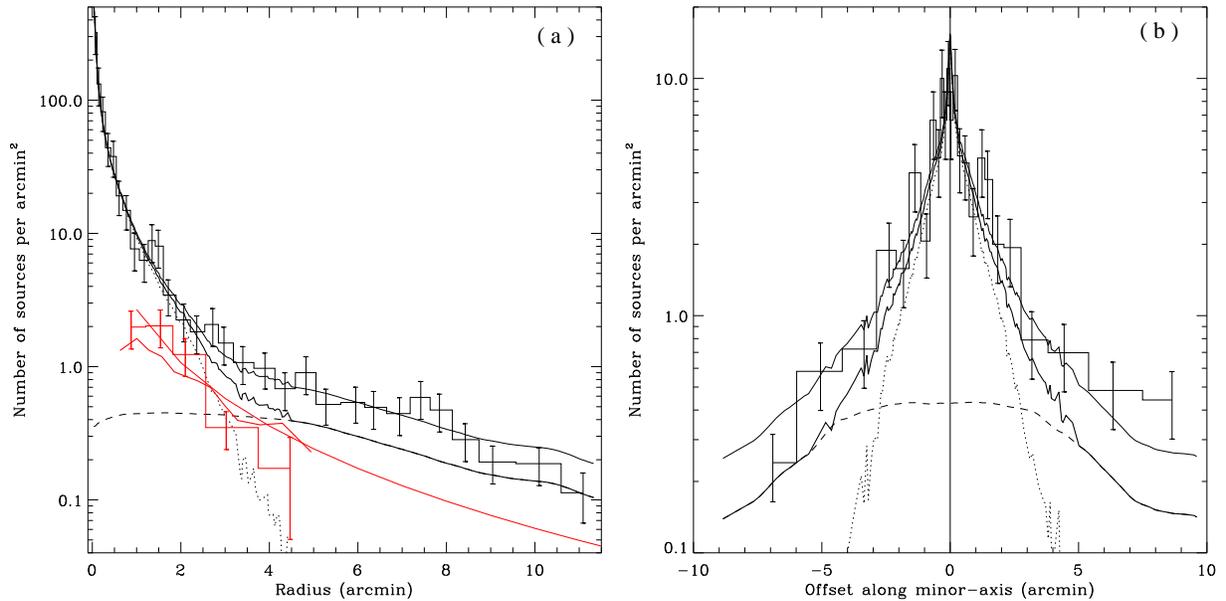,width=\textwidth,angle=0}
   }
\caption{(a) Radial surface density distributions of different classes of sources.
The histograms ({\sl Black}: field sources; {\sl red}: GC-LMXBs) 
are adaptively grouped to have a minimum of 10 sources per bin. 
The thick black solid curve
represents a characterization of the field source distribution, which
consists of contributions of field LMXBs (dotted curve; a normalized
2MASS K-band light radial profile) and a cosmic background of AGNs
(dashed curve). The normalizations of the LMXB and AGN components are 
adopted from Gilfanov (2004) and Moretti et al. (2003), respectively.
The thin black solid curve consists of the same LMXB contribution, but with
1.8 times higher contribution from the AGNs. The thick and thin curves
in red represent 
the distribution of the S06 GCs and a de Vaucouleurs law
characterizing the GC distribution derived by Rhode \& Zepf (2004),
both multiplied by a factor of 0.06. (b) Similar to (a), but for the
vertical surface density distribution of field sources along the galaxy's
minor-axis. The source density is averaged within parallel slices of
10 arcmin in width. North is positive.}
\label{fig:rsb_sou}
\end{figure*}

Fig.~\ref{fig:rsb_sou}a shows the azimuthally-averaged radial
distribution of the source surface
density. The GC-LMXBs cover a radial range between
0\farcm4-4\farcm8, over which they appear 
more concentrated than the entire GC population. This concentration
is likely attributed to  
those sources associated with red GCs (Fig.~\ref{fig:GC})
which have a steeper radial distribution than the blue GCs.
The field sources presumably consist of two components: LMXBs predominantly present in the bulge, and cosmic AGNs dominating the surface density at
large radii. Indeed, the 2MASS K-band starlight (Jarrett et al.~2003), 
when scaled by a factor of 8.1 X-ray sources per $10^{10} L_{\odot,K}$ as
derived for nearby galactic bulges (Gilfanov 2004), is a reasonable
characterization of the surface density distribution of the field sources
within $R\sim$2$^\prime$ (roughly twice the K-band
effective radius of the galaxy), as expected if the field LMXBs are distributed
following the old stellar populations. 

However, at radii outside the optical extent of the galaxy ($R\gtrsim$4$^\prime$), 
deviations in the number of field sources from the expected cosmic contribution are significant. 
The cosmic contribution at a given position across the FoV 
can be determined from 
the Log$N$-Log$S$ relation of cosmic AGNs (Moretti et al.~2003), 
accounting for the local detection threshold (Fig.~\ref{fig:sou_dist}).
Specifically, the detection threshold in terms of F-band count rate is
converted into a 0.5-2 keV flux, as adopted by Moretti et al.~(2003; Eqn.~2 therein),
assuming an intrinsic
power-law spectrum with a photon index of 1.4, suitable for cosmic
AGNs (Moretti et al.~2003),
and the Galactic foreground absorption. 
This cosmic component is shown by a dashed curve in Fig.~\ref{fig:rsb_sou}a.
Between $R=$4$^\prime$-9$^\prime$, the expected number of cosmic AGNs is 52.3,
while the observed number of field sources is 101. 
Only at radii beyond 9$^\prime$ do the observed and predicted surface
density distributions agree with each other (19.2 versus 21; Fig.~\ref{fig:rsb_sou}a).
While the adopted Log$N$-Log$S$ relation is derived from a
combination of shallow wide-field and deep pencil-beam surveys 
(Moretti et al.~2003),
the surface density of cosmic AGNs is expected to vary from field to field. 
The normalized cosmic variance can be estimated as (e.g., Lahav \&
Saslaw 1992)
\begin{equation}
\sigma^2_c = \frac{1}{\Omega^2}\int w(\theta){d\Omega_1}{d\Omega_2} = C_{\gamma}\theta^{\gamma-1}_0\Theta^{1-\gamma}, 
\end{equation}
where $w(\theta)=(\theta/\theta_0)^{1-\gamma}$ is a power-law angular
correlation function (Peebles 1980), $\theta_0$ the correlation length,
$C_{\gamma}$ a numerical factor dependent on the index $\gamma$, and 
$\Omega = \Theta^2~deg^2$ the size of the FoV. For the canonical value
of $\gamma =1.8$ we have $C_{\gamma}\approx 2.25$, and 
we adopt $\theta_0 \approx 0.00214~deg$ (i.e., 7\farcs7) measured from
a serendipitous {\sl XMM-Newton} survey reaching a flux limit of
$F_{0.5-2 {\rm~keV}} \sim 10^{-15} {\rm~ergs~s^{-1}~cm^{-2}}$ (Ebrero
  et al.~2009), comparable to the flux limit achieved in the Sombrero field. 
Hence the fractional cosmic variance is 
$\sigma_c \approx 0.16$ for the $4^\prime-9^\prime$ annulus in
which the source overdensity is found. Combining the cosmic and
Poisson variances ($\sigma=\sqrt{\sigma_c^2+\sigma_P^2} \approx 0.21$), the significance
of the overdensity is $\sim$4.4~$\sigma$.
Unidentified Galactic interlopers may contribute to the 
overdensity, but this
is unlikely given the relatively high Galactic latitude of
Sombrero. On the other hand, a small fraction of the identified interlopers can in fact be
background galaxies or AGNs. The removal of such sources has the effect of  
reducing the local cosmic background, and hence the actual overdensity is likely
more significant than estimated here. 
Neither can the overdensity be accounted for by
simply increasing the normalization of the galactic bulge component, as the
K-band surface brightness drops steeply with radius and becomes negligible beyond
$\sim$$4^\prime$.
 We discuss in \S~\ref{sec:dis} the possibility that the
overdensity originates from sources physically associated with the halo of Sombrero.

\subsection{Number-flux relation} 
\begin{figure*}[!htb]
\centerline{
  \epsfig{figure=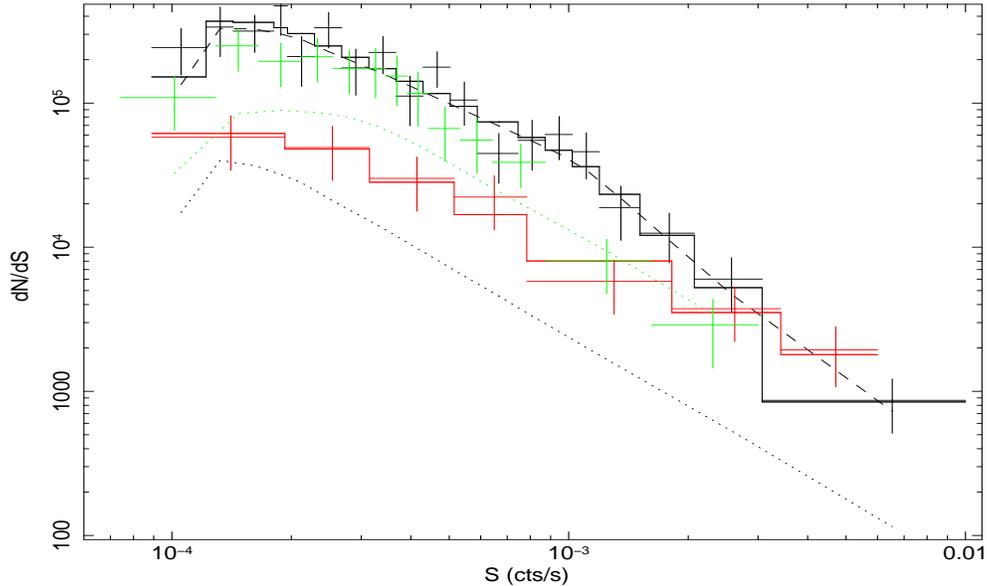,width=0.8\textwidth,angle=0}
   }
\caption{Differential number-flux relation of field sources
  detected between 0\farcm1-4$^\prime$ (A1; {\sl black}) and
  4$^\prime$-9$^\prime$ (A2; {\sl green}), and sources associated with GCs
  ({\sl red}). The count rates are measured in the F-band (0.4-6 keV) for A1
  sources and GC-LMXBs and in the M-band (0.7-2 keV) for A2 sources.
Data points 
 are adaptively grouped to have a minimum of 6 sources per bin. 
The black histogram represents an acceptable fit to the A1 NFR,
consisting with a broken power-law for the galactic component (dashed
curve) and a cosmic component from Moretti et al. (dotted curve).
The red histogram represents an acceptable power-law fit to the GC NFR.
The green dotted curve represents the expected cosmic component in the
A2 NFR. Note the turnover of the NFRs and models at low count rates, 
which results from the incompleteness and Eddington bias in the source
detection.}
\label{fig:LF}
\end{figure*}

{\label{subsec:nfr}}
We construct differential number-flux relations (NFRs; Fig.~\ref{fig:LF}) for the
field sources detected in the two annuli with inner-to-outer radii of
10$^{\prime\prime}$-4$^\prime$ (hereafter A1) and 4$^\prime$-9$^\prime$
(hereafter A2) and for the GC-LMXBs. 
The inner radius of A1 is chosen to minimize the uncertainty due to
source confusion in the 
innermost regions. We also exclude a region approximately coincident
with the dust lane (enclosed by the dashed ellipses in
Fig.~\ref{fig:sou_plot}), where both HMXBs and additional interstellar 
absorption 
are likely present. Moreover, we consider only
sources detected in the F-band for NFR A1 and
only sources detected in the M-band (0.7-2 keV) for NFR A2. 
The latter, in particular, ensures an optimal comparison with the 0.5-2 keV
Log$N$-Log$S$ relation of Moretti et al.~(2003). 
A1 and A2 contain 159 and 93 sources, respectively.
As the characteristic spectra of LMXBs and cosmic AGNs are not
identical, we choose to refer ``flux'' to the observed count rate,
instead of a converted incident flux.

Our analysis of the NFRs follows the procedure described in Wang (2004), 
accounting for the dependence of the detection completeness on the local PSF,
effective exposure and background, as well as the so-called X-ray Eddington bias that describes the probability
distribution of observed count rates due to the Poisson uncertainties
and the intrinsic slope of the NFR. The 90\% completeness limit is 
estimated to be $\sim$$2\times10^{37}{\rm~ergs~s^{-1}}$ for sources
detected in A1 and the GC-LMXBs, while this limit is significantly higher
in A2 ($\sim$$6\times10^{37}{\rm~ergs~s^{-1}}$).
As sources in A1 are dominated by the galactic old
stellar populations,
we use the K-band starlight distribution as a spatial weight to
calculate the accumulated functions of the incompleteness and
Eddington bias for these galactic sources, whereas the intrinsic distribution
of the cosmic AGNs is assumed to be uniform across the field.
While approximately half of the sources in A2 are possibly
associated with Sombrero (\S~\ref{subsec:spatial}), we also assume a uniform spatial distribution for
these sources, for simplicity.

As for the radial surface density distributions (\S~\ref{subsec:spatial}), 
we fit NFR A1 with two components: a cosmic component, 
assuming the Log$N$-Log$S$ relation of Moretti et al.~(2003), and a galactic 
component. The cosmic component contributes $\sim$10\% to the NFR. We begin with a power-law model for the galactic component,
\begin{equation}
\left({dN \over dS}\right)= K S^{-\alpha},
\label{eq:nfr}
\end{equation}
where $S$ is in units of ${\rm cts~s^{-1}}$. The fit gives $\alpha =
1.59^{+0.08}_{-0.08}$ and $K = 0.44^{+0.37}_{-0.20}$ sources per ${\rm
  cts~s^{-1}}$. 
With C-statistic/d.o.f. = 22.1/17, this fit is only acceptable at a
confidence level of 29\%. A broken power-law
(Fig.~\ref{fig:LF}) gives a much improved fit with 
C-statistic/d.o.f. = 11.4/15, acceptable at a confidence level of 91\%. The fitted power-law indices
are $1.23^{+0.13}_{-0.14}$ and $2.43^{+0.31}_{-0.27}$ below and above the break of
$1.05^{+0.42}_{-0.24}\times 10^{-3} {\rm~counts~s^{-1}}$, which
corresponds to a 0.5-8 keV intrinsic luminosity of $1.0\times 10^{38}{\rm~ergs~s^{-1}}$.
 The change in the slope is of $\sim$ 4$\sigma$ significance; the fitted indices 
are consistent with those derived from deep {\sl Chandra}
observations of three elliptical galaxies NGC 3379, NGC 4278 and NGC
4697 (Kim et al.~2009).    
The break luminosity, on the other hand, is marginally higher than
that found in the elliptical galaxies ($\sim$$6\times 10^{37}{\rm~ergs~s^{-1}}$).

For the NFR of GC-LMXBs, we consider only the galactic component, for which a
power-law model, with an index of $1.13^{+0.14}_{-0.14}$ and
C-statistic/d.o.f. = 1.2/5,  
provides an acceptable fit at a confidence level of 99\%. 
Due to the limited number statistics, there is no obvious need
for a broken power-law model to characterize this NFR.
Finally, we find that the Log$N$-Log$S$ relation of Moretti et
al.~(2003) is inconsistent with the NFR A2 (C-statistic/d.o.f. = 46.7/9), in particular falling
short at M-band count rates below $9\times10^{-4} {\rm~cts~s^{-1}}$
(Fig.~\ref{fig:LF}). 
Instead, the NFR A2 can be well fitted by a power-law model with a
slope of $2.03^{+0.12}_{-0.12}$, significantly steeper than 
Moretti et al.'s slope of $\sim$1.60 over the considered flux range.  
The removal of background interlopers (\S~\ref{subsec:nfr}) may steepen the NFR A2 at the bright end. To account for this effect, we reconstruct the NFR A2 without removing any identified interlopers. This new NFR can be fitted by a power-law with a slope of $1.89^{+0.12}_{-0.12}$.   
This again suggests that sources detected in A2 
originate in part from a population distinct from cosmic AGNs.  

\subsection{Unresolved X-ray emission from GCs} 
\begin{figure*}[!htb]
\centerline{
  \epsfig{figure=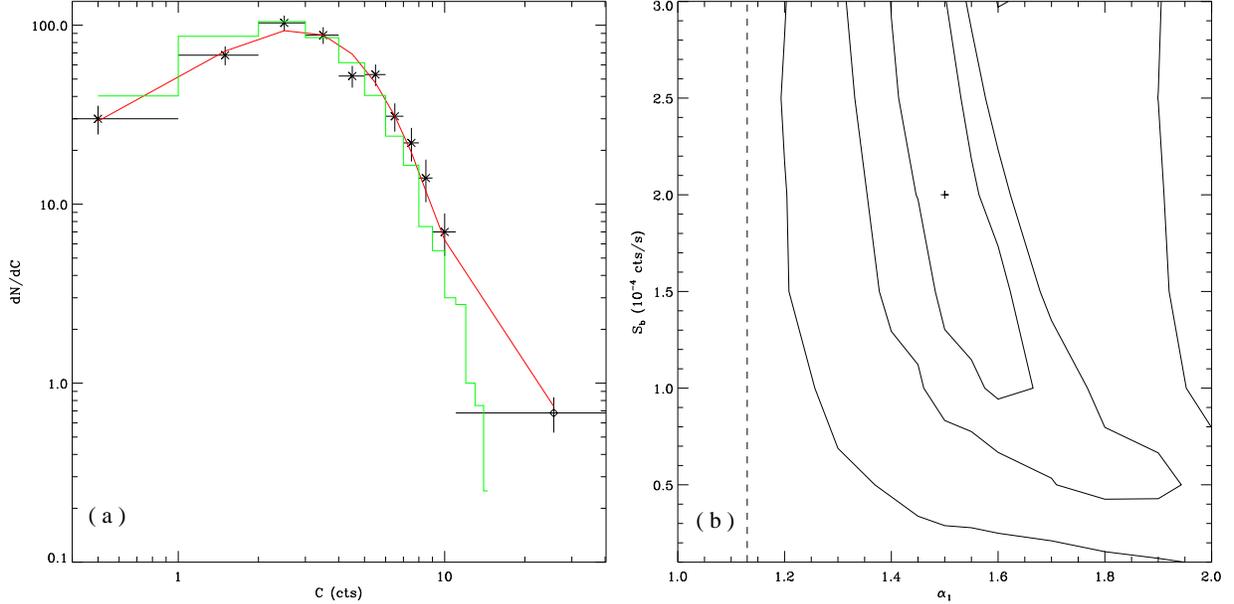,width=\textwidth,angle=0} 
}
\caption{(a) Number distribution of source counts for the GCs: {\sl
  crosses}: X-ray-undetected GCs; {\sl diamond}: X-ray-detected GCs
  with fluxes below the detection completeness limit. The histogram
  is a representation of the number distribution of local background counts. 
  The curve shows the predicted number distribution from the best-fit 
  model NFR. See text for details. (b) The 68\%, 90\% and 99\%
  confidence contours of the faint-end index and the break flux. The
  plus sign marks the best-fit values. The vertical dash line marks
  the best-fit value of the bright-end index derived from the
  X-ray-detected GCs.}
\label{fig:GCanal}
\end{figure*}

{\label{subsec:GCunr}}
While 41 GC-LMXBs are found (\S~\ref{subsec:GC}), the
rest (93.8\%) of the S06 GCs remain undetected in X-rays. 
Nevertheless, some X-ray emission is expected to come from these GCs. 
For instance, some GCs may host an LMXB that 
is fainter than the local detection threshold
($\sim$$10^{37}{\rm~ergs~s^{-1}}$). 
The ``X-ray emission'' from the undetected GCs can  be
quantified through a {\sl fluctuation analysis} (e.g., Miyaji \&
Griffiths 2002; Hickox \& Markevitch~2007; Hickox et al.~2009) and subsequently
provides useful constraints on the GC-LMXB population, such as the
shape of the NFR at fluxes significantly below the detection
threshold. The procedure is as follows.  
First, we collect 0.4-6 keV source counts registered within the 90\% EER
(ranging from 1\farcs5-5\farcs5)
around individual undetected GCs.    
To minimize contamination from nearby X-ray sources, a GC is
excluded if it is located within three times the 90\% EER of a detected
X-ray source. This results in a total of 483 GCs, the source counts of
which range from 0 to 18 cts. 
The number distribution of source counts is plotted in Fig.~\ref{fig:GCanal}a.
showing a bump representing the statistical fluctuation of the local
background and a high-count tail presumably arising from the collective emission from the GCs. 
For comparison, we also show the mean number distribution of counts
collected in a similar way from ``empty'' regions (histogram in
Fig.~\ref{fig:GCanal}a), i.e, positions $\pm5^{\prime\prime}$ in
R.A. and Dec. from the GC centroids.
The distribution of counts collected from the GCs shows a clear excess 
above 5 cts. Corrected for the mean exposure at the GC positions,
this excess corresponds to an integrated net count rate of
$(2.1\pm0.1)\times10^{-3}{\rm~cts~s^{-1}}$, or 7.4 $\sigma$ over the local background.
Repeating this exercise on a subset of the GCs, we estimate an integrated net count rate of 
$(7.5\pm0.7)\times10^{-4}{\rm~cts~s^{-1}}$ for 101 red GCs brighter than the turnover magnitude,
$(5.3\pm0.6)\times10^{-4}{\rm~cts~s^{-1}}$ for 149 blue GCs brighter than the turnover magnitude,
$(6.9\pm0.7)\times10^{-4}{\rm~cts~s^{-1}}$ for 116 red GCs are fainter than the turnover magnitude,
$(1.1\pm0.3)\times10^{-4}{\rm~cts~s^{-1}}$ for 117 blue GCs are fainter than the turnover magnitude,
respectively. The red GCs apparently also show a higher integrated unresolved X-ray flux than the blue GCs.

Second, we statistically constrain the net contribution from the
undetected GCs, and moreover, their NFR at low fluxes, in 
the following steps. 

i) The shape of the NFR is assumed to be a broken power-law:
\begin{eqnarray}
     \frac{dN_{\rm GC}}{dS}=
     \left\{ \begin{array}{l@{\quad \quad}l}
       K S^{-\alpha_1} &
                (S_{\rm min}\leq S < S_{\rm b}), \\
       K S_{\rm b}^{\alpha_2-\alpha_1}S^{-\alpha_2} &
                (S_{\rm b}\leq S), \\
                \end{array}
     \right.
\label{eq:bknpow}
\end{eqnarray}
where $S_{\rm min}=10^{-6}{\rm~cts~s^{-1}}$ is the minimum flux
above which the NFR is evaluated. This value
corresponds to a luminosity of $\sim$$10^{35}{\rm~ergs~s^{-1}}$,
reasonable for LMXBs. We note that
a fluctuation analysis is typically sensitive to 1 $\sigma$ below the mean
background ($\sim$2 cts here), which is $\sim$0.6 cts, corresponding to a count rate of $\sim$3$\times10^{-6}{\rm~cts~s^{-1}}$.
We fix the bright-end index,
$\alpha_2$, at the value of 1.13, justified
by the goodness of the power-law characterization for the
identified GC-LMXBs in \S~\ref{subsec:nfr}.
The normalization $K$ is effectively determined by the total number of
detected GC-LMXBs with count rates above the 100\%
completeness limit of $3\times10^{-4}{\rm~cts~s^{-1}}$ (28 of 659 GCs).  
Therefore the free parameters are the faint-end index, $\alpha_1$, and
the break flux, $Sb$, which is assumed to be no higher than $3\times10^{-4}{\rm~cts~s^{-1}}$.

ii) With Monte Carlo simulations, we generate source counts from the 483
undetected GCs, based on the above NFR. The expected source count of a GC is
the sum of the local background flux (which has been determined in the
source detection procedure) and the NFR-predicted flux,
multiplied by the local effective exposure and then
Poisson-randomized. In each simulation run, a distribution of
$N_{{\rm sim},i}$, the number of GCs with $i$
counts, is generated. The deviation between the simulated and observed distribution
(Fig.~\ref{fig:GCanal}) is evaluated by the modified C-statistics
(Cash 1979), 
\begin{equation}
C = 2 \sum_i [N_{{\rm sim},i}-N_{{\rm obs},i}+N_{{\rm obs},i}({\rm
  ln}N_{{\rm sim},i}-{\rm ln}N_{{\rm obs},i})].
\end{equation}
    
iii) For a given pair of $\alpha_1$ and $S_b$, 1000 simulations are run and the
mean value of $C$ is calculated. 
A best-fit is found at
the minimum of $C/d.o.f.$=7.22/10, which is acceptable at a confidence
level of 97\%. The fit is not sensitive to $S_b$, but gives  
$\alpha_1=1.55^{+0.15}_{-0.15}$ (Fig.~\ref{fig:GCanal}b). Comparing with the bright-end index
$\alpha_2=1.13^{+0.14}_{-0.14}$, this rules out a flattened NFR toward
fainter fluxes at a 2 $\sigma$ significance. 
The corresponding mean value of $N_{{\rm obs},i}$
is shown in Fig.~\ref{fig:GCanal}a. 

iv) The above procedure has two implicit assumptions: (1)
the source counts from individual GCs are independent, and (2) the 
GC X-ray flux is not dependent on other GC properties. 
While the first assumption is generally true given the small sky area occupied
by the GCs, the second assumption is inconsistent with the trend that
more luminous and denser GCs have a larger chance to host an LMXB.
To test this effect, we repeat our procedure for
those GCs above the turn-over magnitude, so that 
the dependency on GC properties is minimized, but this comes
at the price of reduced statistics. The corresponding best-fit $\alpha_1$ 
is $1.42^{+0.26}_{-0.19}$, again implying that the NFR does not
flatten toward fainter fluxes. 

v) Stellar populations other than LMXBs, such as cataclysmic variables
(CVs) and millisecond pulsars, can also contribute to the detected
X-ray counts in individual GCs. The collective luminosity of such populations
correlates with GC mass, and is expected to be
$\lesssim10^{35}{\rm~ergs~s^{-1}}$ per GC. To evaluate the effect of such a
contribution, we simply add a delta
function at $10^{-6}{\rm~cts~s^{-1}}$ to the NFR model and repeat the simulations. This results
in a best-fit $\alpha_1$=$1.40^{+0.20}_{-0.16}$, again showing no evidence of a 
flattened NFR at faint fluxes.

\subsection{Variability} \label{subsec:var}
\begin{figure*}[!htb]
\centerline{
 \epsfig{figure=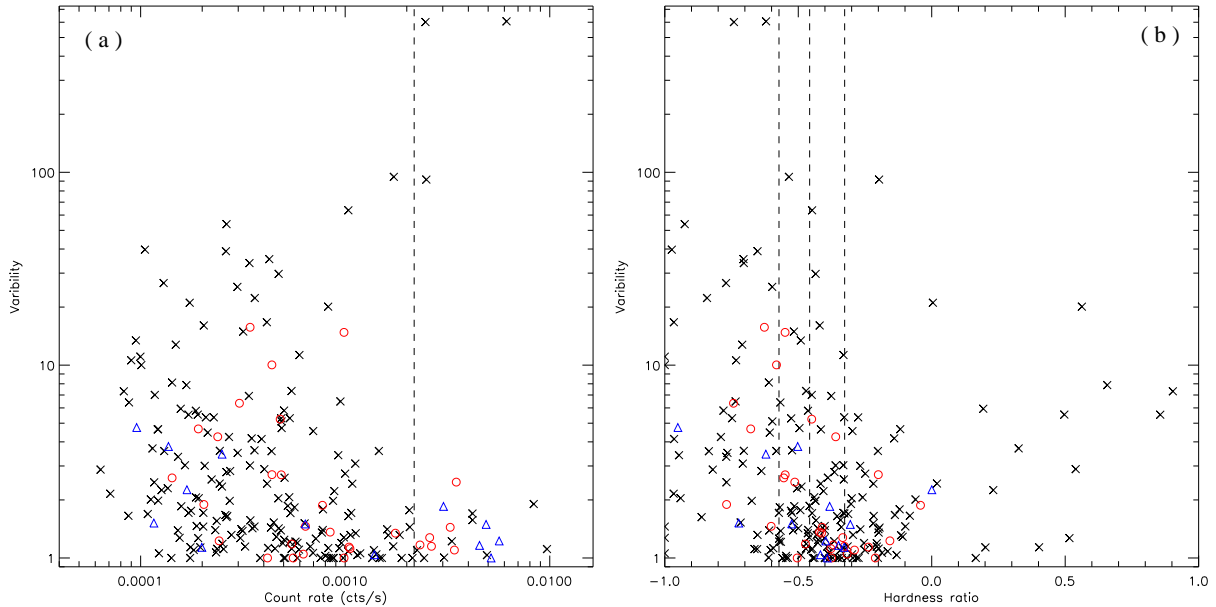,width=\textwidth,angle=0}
 }
\caption{Source variability (see text for definition) vs. (a) count
  rate and (b) hardness ratio. Field sources detected within
  $R=4^\prime$  are shown by crosses,
  while LMXBs detected in blue and red GCs are shown by blue triangles and red
  circles, respectively.
The dashed line in (a) represents the Eddington luminosity of an
accreting NS with a mass of 1.4 M$_{\odot}$.
The dashed lines in (b), from left to right, correspond to hardness ratios 
of an absorbed power-law spectrum with a fixed $N_{\rm H}=10^{21}{\rm~cm^{-2}}$ and a photon index of 2.0, 1.7 and 1.4, respectively.}
\label{fig:vr}
\end{figure*}

Transients and variable X-ray sources are often found in early-type
galaxies. Here we quantify the variability of the
GC-LMXBs and the field sources detected within $R=4^\prime$, where the
FoV is common to all three observations. The net count rate of each 
source in each of the three observations is determined in a
homogeneous way, similar to the determination of the average source 
count rate from the combined data.
If a source is below the detection threshold in a
given observation, its net count
rate is measured in the same way as described in \S~\ref{subsec:GCunr}.
We define source variability $V=F_{\rm h}/F_{l}$, where $F_{\rm h}$ is the
highest detected count rate among individual detections, and $F_{\rm l}$
the statistical upper limit of the lowest detected count rate. 
The variability is plotted versus the
average count rate (Fig.~\ref{fig:vr}a) and the hardness ratio of individual sources (Fig.~\ref{fig:vr}b).
$\sim$44\% (83 of 188) of the field sources and $\sim$36\% (15 of 41)
of the GC-LMXBs exhibit $V>2$.
$\sim$13\% (24 of 188) of the field sources are strongly
variable (defined as $V>10$), whereas this number is 
$\sim$7\% (3 of 41) in GC-LMXBs. 
Two red and one blue GC-LMXBs are detected in only one observation, but none of them has  $V>10$.
We also find that 53 of the field sources are detected in only one observation, and thus may be transient sources. 17 of these 53 sources, or $\sim$9\% of the total field sources considered, exhibit $V>10$ 
and hence are the most probable transients.
For comparison, five and three transients, also defined as $V>10$, are found among
$\sim$100 source detected in NGC3379 and $\sim$180 sources in NGC4278, 
respectively (Brassington et al.~2008, 2009). In another elliptical galaxy, 
NGC~4636, which is comparable to Sombrero in stellar mass but more massive than 
NGC3379 and NGC4278, two such transient candidates are found among 
$\sim$230 detected sources (J. Posson-Brown, private communication).

The two most variable sources (with $V>600$) exhibit
a super-Eddington luminosity for an NS, as well as a hardness ratio
softer than typical NSs, implying that they are likely accreting BHs. 
A close examination of the data reveals that the brighter one of 
these two sources, centered at 
[R.A.,DEC.]=[12:40:01.85, -11:36:15.3], is only visible in 
Obsid.~9532, and the fainter one, centered at 
[R.A.,DEC.]=[12:40:00.95, -11:36:54.1], only appears in Obsid.~1586.  
On the other hand, the super-Eddington GC-LMXBs show no significant
variability, suggesting that they are superpositions of 
accreting objects, most likely NSs.

\section{Discussion} {\label{sec:dis}}
\subsection{LMXB luminosity function}
In \S~\ref{subsec:GCunr} we quantify
the collective X-ray
emission from GCs below the X-ray detection threshold. In particular,
the number-flux relation of these sources is examined. Hereafter we
use the term ``luminosity function'' (LF) for formality. Our
fluctuation analysis rules out a slope of the LF flatter than 1.1 below  
$10^{37}{\rm~ergs~s^{-1}}$ at 2 $\sigma$ significance. In fact, the inferred faint-end power-law index
is steeper than that of the bright-end. This is contrary to the
findings in recent X-ray studies of GC populations in several elliptical galaxies,
including NGC 3379, NGC 4278, NGC 4697 (Kim et al.~2009) and NGC 5128
(Voss et al.~2009), and in the bulge of M31 (Voss \& Gilfanov 2007), in which deep {\sl Chandra} observations allow for
detection of faint GC-LMXBs down to $10^{36}-10^{37}
{\rm~ergs~s^{-1}}$. In these galaxies the GC LF appears flattened at
the faint-end with
respect to its slope at the bright-end. 

\begin{figure*}[!htb]
\centerline{
     \epsfig{figure=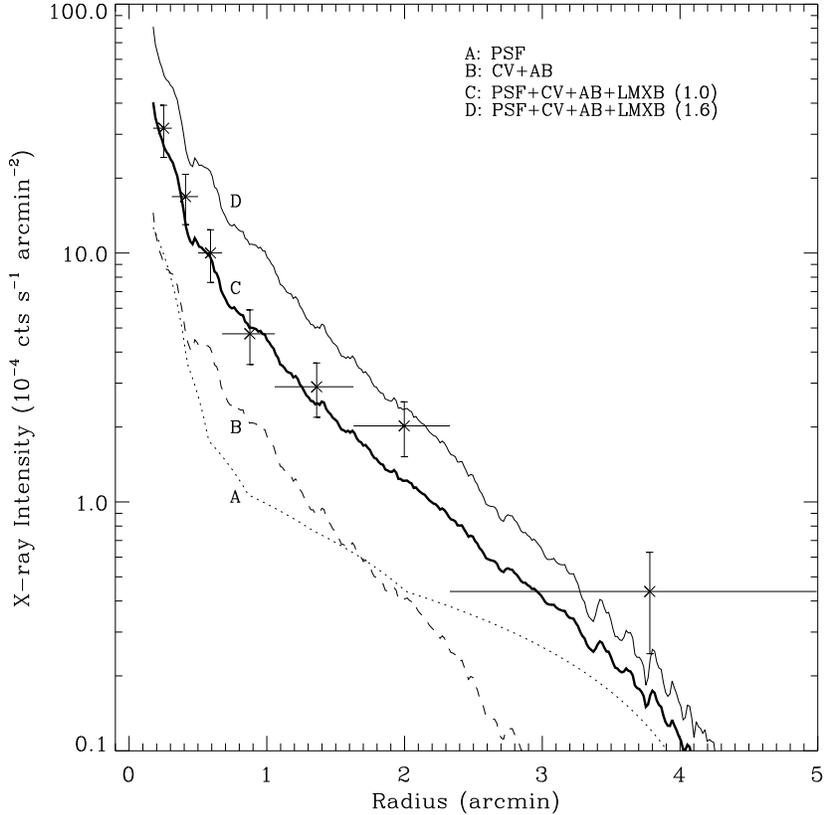,width=0.7\textwidth,angle=0}
  }
\caption{Background-subtracted, exposure-corrected radial intensity
  distributions of the unresolved 2-6 keV emission, shown by {\sl
  crosses}. The estimated contribution to the
  unresolved emission from PSF scattering of detected sources (dotted
  curve) and CV+AB (dashed curve) are marked. The two solid curves
  represent the sum of PSF+CV+AB+LMXB (unresolved), the latter estimated by
  assuming assuming a LMXB LF, the faint-end power-law index of which 
is either 1.0 or 1.6. See text for details.}
\label{fig:unr}
\end{figure*}

The LF of field LMXBs determined in \S~\ref{subsec:nfr}, on the
other hand, is consistent with previous studies in which a single power-law
index of $\sim$1.8 is typically found. We can also examine its
behavior below our detection limit, utilizing the unresolved X-ray emission.
In particular, the 2-6 keV unresolved emission is thought to almost
entirely originate from stellar populations consisting of faint,
unresolved LMXBs and even fainter stellar objects, such as coronally
active binaries (ABs) and CVs (Sazonov et
al.~2006; Revnivtsev et al.~2006, 2007, 2008). Fig.~\ref{fig:unr}
shows the azimuthally-averaged radial intensity distribution of the 2-6 keV 
unresolved emission (crosses), after removal of all detected sources
with F-band count rates above the completeness limit of
$3\times10^{-4}{\rm~cts~s^{-1}}$. 
 Our source-removal procedure effectively
excludes $\sim$96\% of the source photons. The 4\% PSF-scattered photons can be
accounted for by following the radial distribution of detected bright sources.
This PSF-scattered component is shown in Fig.~\ref{fig:unr} by  
a dotted curve.
In bulges and elliptical galaxies, the collective 2-6 keV emissivity (per stellar
mass) of CV+AB is likely universal and has been calibrated to better
than $\sim$15\% (Revnivtsev et al.~2008). The CV+AB contribution 
for Sombrero is shown in Fig.~\ref{fig:unr} by a normalized
K-band radial intensity distribution (dashed curve).
The contribution of faint LMXBs to the 2-6 keV unresolved emission is
then determined by extrapolating a given LF. We test different values
of the index $\alpha$ for LMXBs toward the faint-end, fixing the bright-end index at 1.6
(\S~\ref{subsec:nfr}) and the break luminosity at the completeness limit. The sum of PSF+(CV+AB)+LMXB is shown by solid
curves in Fig.~\ref{fig:unr} for representative values of $\alpha$=1.6
(no flattening) and $\alpha$=1.0. Clearly the observed unresolved
emission indicates a flattened LF for the field LMXBs below $\sim$$3\times10^{37}{\rm~ergs~s^{-1}}$.

\subsection{Origin of the source overdensity}
In \S~\ref{subsec:spatial} we show an excess of
X-ray sources detected at large
galactocentric radii with respect to
the expected number and typical spatial variance of cosmic AGNs. 
These sources are plausibly
associated with Sombrero. Here we discuss 
their possible parent population.

In light of the presence of a large GC population in Sombrero to a
galactocentric radius of $\sim$50 kpc (Rhode \& Zepf 2004), a possible
host of these X-ray sources is GCs located at large radii. Based on ground-based
observations, Rhode \& Zepf (2004) reported a
total of $\sim$1900 GC candidates, whose surface density distribution is well characterized by a de Vaucouleurs law with an effective
radius of $\sim$6\farcm2, as shown in
Fig.~\ref{fig:rsb_sou}. This surface
density distribution predicts that $\sim$22\% of the whole GC population
is located between $R=$4$^\prime$-9$^\prime$, where the X-ray source
overdensity is found. Assuming that 6\% of these GCs host a detectable
X-ray source, a value appropriate for the S06 GCs, 32.5
GC-LMXBs are expected, which is marginally sufficient to account
for the observed
overdensity. However, it is premature to claim that 
GCs are the primary host
of the observed overdensity, for the following reasons: 1) while there are 96
S06 GCs identified at $R\geq4^\prime$, only 3 of them ($\sim$3\%) are found 
to be associated with an X-ray source. The higher percentage (6\%) of associations
for the entire S06 GCs presumably arises from
associations with red GCs which have a steeper concentration in the bulge
than the blue ones; 2) at large radii, contamination of interlopers
becomes increasingly
significant for the Rhode \& Zepf GC candidates and the true GC
population may have a steeper surface density distribution. Indeed, in
a spectroscopic follow-up study,
Bridges et al.~(2007) found a substantial fraction of the Rhode \&
Zepf GC candidates to be interlopers;  3) The
same GC distribution of Rhode \& Zepf also predicts $\lesssim$9.8 X-ray sources detectable (with a detection threshold of
$5\times10^{-4}{\rm~cts~s^{-1}}$, or 
$\sim$$5\times10^{37}{\rm~ergs~s^{-1}}$) at $R=9^\prime$-11\farcm5, 
or a mean surface density of $\lesssim$0.06
source per arcmin$^2$ over the cosmic component, which is not observed.
The Rhode \& Zepf (2004) 
GCs, whose sky positions are unpublished, could be useful to directly test
this possibility.

Another possibility is binary systems, favorably
containing an accreting NS, which are ejected from the inner galactic
regions due to the recoil of the system after the supernova that created
the NS. In Fig.~\ref{fig:rsb_sou}b the surface density distribution of
field sources along the minor-axis of the disk is shown. Compared to
the starlight distribution (dotted curve in Fig.~\ref{fig:rsb_sou}b), the detected
sources show a clear overdensity as close as 1 arcmin from the midplane. This 
is reminiscent of a distribution of sources ejected from the disk.
The spatial distribution of Galactic NSs resulting from
supernova kicks has long been the subject of studies, e.g., by 
Paczy$\acute{n}$ski (1990), among others. Recently, Zuo, Li \& Liu (2008)
modelled the spatial distribution of Galactic X-ray binaries,
accounting for the kinematic evolution of the kicked binary systems.
However to our knowledge, a similar study in the scope of
galactic spheroids (i.e., bulges of early-type spirals and 
elliptical galaxies) has not been carried out. 

A third and perhaps more
controversial possibility is X-ray binaries formed in relaxed remnants of recent mergers,
which are now falling back to the inner galactic regions, as suggested
by Zezas et al.~(2003) for the field sources detected in two
elliptical galaxies, NGC4261 and NGC4697. These sources show 
an azimuthally non-uniform distribution toward large radii. 
We have examined the azimuthal distribution of the sources detected 
with $R=$4$^\prime$-9$^\prime$, but found no significant non-uniformity.

\begin{figure*}[!htb]
\centerline{
      \epsfig{figure=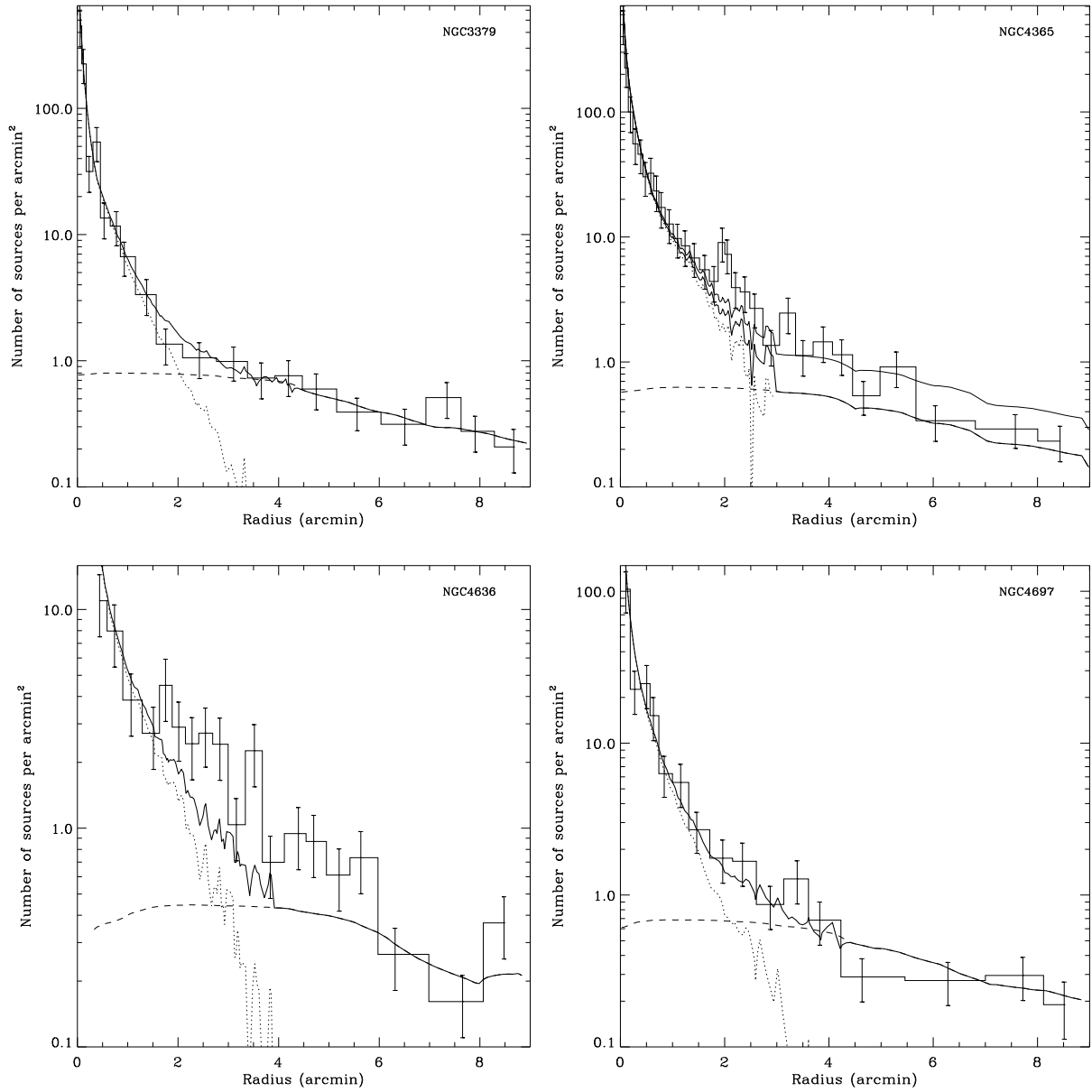,width=\textwidth,angle=0}
       }
\caption{Radial surface density distributions for sources detected in
  (a) NGC 3379, (b) NGC 4365, (c) NGC 4636 and (d) NGC4697. The curves
have the same representation as in Fig.~\ref{fig:rsb_sou}.}
\label{fig:Es}
\end{figure*}

It is conceivable that a similar source overdensity also exists in
typically massive elliptical galaxies. We search the {\sl Chandra}
archival data for suitable galaxies to test this possibility.
Four elliptical galaxies are thus selected, including NGC 3379,
NGC 4365, NGC 4636 and NGC 4697. The distances (10-20 Mpc) and cumulative
{\sl Chandra} exposure ($\sim$200 ks) of these galaxies allow for 
a similar source detection threshold and a similar coverage
of radial extent as we achieved for
Sombrero. Indeed, the discrete source populations in each of these
galaxies have been the subject of recent studies (Brassington et
al.~2008; Sivakoff et al.~in preparation; Posson-Brown et al.~2009;
Sivakoff et al.~2008). Here we perform source detection 
and construct a radial source density profile for each galaxy (Fig.~\ref{fig:Es}), 
in the same way as for Sombrero, except that no GC-LMXBs are
identified due to the lack of a published GC catalogue for any of
these galaxies. As shown in Fig.~\ref{fig:Es}, 
the surface density profiles of NGC 3379 and NGC 4697 can be described
 by a galactic component following the K-band starlight
and a cosmic component,
whereas the other two galaxies, NGC 4365
and NGC 4636, exhibit a source overdensity above the expected cosmic
contribution in regions beyond their half-light radii. NGC 4365 and NGC 4636 are two massive Virgo
ellipticals, with a total stellar mass even higher than that of Sombrero,
whereas NGC 3379 and NGC 4697 are each about 3 times less massive 
than Sombrero. We emphasize, however, that unidentified GC-LMXBs must 
contribute to part, if not all, of the observed overdensity in NGC
4365, NGC 4636 and Sombrero.

\section{Summary} {\label{sec:sum}}
Our study of the X-ray sources in the Sombrero galaxy can
be summarized as follows:

1. With a detection limit of $10^{37}{\rm~ergs~s^{-1}}$, a total of
   383 sources are detected within a projected galactocentric radius of
   $R$=11\farcm5. Among them, 41 sources, presumably LMXBs, are found
   to be associated with GC candidates identified through
   {\sl HST} observations covering the optical extent of Sombrero ($R\sim$4). 
   28 of the 41 sources are found in metal-rich GCs, indicating that
   metal-rich GCs have a higher probability of hosting an LMXB. On the other
   hand, metal-poor GCs host the four
   brightest GC-LMXBs whose individual luminosities exceed the
   Eddington luminosity of an accreting NS.

2. The differential number-flux relation (i.e., luminosity function)
   of the detected GC and field sources are quantified by a power-law
   model. The slopes are $\sim$1.1 and $\sim$1.6 for the GC and field
   sources, respectively, consistent with previous findings in nearby
   elliptical galaxies. 

3. Photon counts from the positions of X-ray-undetected GCs show a
   $\sim$7.4 $\sigma$ excess above the local background. Based on these
   counts, a fluctuation
   analysis shows that the differential number-flux relation
   does not flatten at fluxes below $\sim$$10^{37}{\rm~ergs~s^{-1}}$, contrary
   to recent findings in several elliptical galaxies and the bulge of M31.

4. For field sources, the 2-6 keV unresolved emission places a tight
   constraint on the differential number-flux relation, implying a flattened slope of $\sim$1.0 below
   $\sim$$10^{37}{\rm~ergs~s^{-1}}$.

5. A total of 101 sources are detected in the halo of Sombrero. This
   is a $\sim$4.4 $\sigma$ excess above the expected number of cosmic AGNs,
   indicating that either the cosmic background is unusually high in this direction or about half of these sources are associated with Sombrero.  

\vspace{1 cm}
We are grateful to Ryan Hickox for his advice on the fluctuation analysis.
This work is supported by the SAO grant G08-9088.

\clearpage

\begin{deluxetable}{cccccccccccc}
\centering
\rotate
  \tabletypesize{\scriptsize}
  \tablecaption{{\sl Chandra} Source List \label{acis_source_list}}
  \tablewidth{0pt}
  \tablecolumns{12}
  \tablehead{
  \colhead{Source} &
  \colhead{CXOU Name} &
  \colhead{$\delta_x$ ($''$)} &
  \colhead{CR}  &
  \colhead{CR1} &
  \colhead{CR2} &
  \colhead{CR3} &
  \colhead{CR4} &
  \colhead{HR} &
  \colhead{HR1} &
  \colhead{HR2} &
  \colhead{Flag} \\
  \noalign{\smallskip}
  \colhead{(1)} &
  \colhead{(2)} &
  \colhead{(3)} &
  \colhead{(4)} &
  \colhead{(5)} &
  \colhead{(6)} &
  \colhead{(7)} &
  \colhead{(8)} &
  \colhead{(9)} &
  \colhead{(10)} &
  \colhead{(11)} &
  \colhead{(12)} 
  }
 \startdata
   1 &  J123919.62-113214.9 &  1.0 &$     1.76  \pm   0.24$  &$    -0.07  \pm   0.07$  &$     0.34  \pm   0.12$  &$     0.94  \pm   0.15$  &$     0.55  \pm   0.14$  &$    -0.38  \pm  -0.01$  &$     0.55  \pm   0.01$  &$    -0.26  \pm  -0.01$  &B, M, H2    \\
   2 &  J123920.06-113628.5 &  1.4 &$     0.60  \pm   0.17$  &$     0.18  \pm   0.11$  &$     0.12  \pm   0.06$  &$     0.26  \pm   0.08$  &$     0.04  \pm   0.07$  &$    -0.88  \pm  -0.04$  &$    -0.10  \pm  -0.03$  &$    -0.76  \pm  -0.09$  &B, M        \\
   3 &  J123920.46-113258.5 &  2.1 &$     0.31  \pm   0.13$  &$    -0.07  \pm   0.00$  &$     0.05  \pm   0.06$  &$     0.36  \pm   0.10$  &$    -0.03  \pm   0.07$  &$    -0.96  \pm  -0.02$  &$     0.96  \pm   0.02$  &$    -0.96  \pm  -0.02$  &M           \\
   4 &  J123920.59-113915.6 &  1.2 &$     0.58  \pm   0.15$  &$    -0.02  \pm   0.06$  &$     0.03  \pm   0.04$  &$     0.50  \pm   0.11$  &$     0.08  \pm   0.07$  &$    -0.72  \pm  -0.01$  &$     0.97  \pm   0.01$  &$    -0.72  \pm  -0.01$  &B, M        \\
   5 &  J123920.69-113408.1 &  1.3 &$     0.81  \pm   0.17$  &$    -0.01  \pm   0.06$  &$     0.18  \pm   0.08$  &$     0.42  \pm   0.10$  &$     0.22  \pm   0.09$  &$    -0.46  \pm  -0.01$  &$     0.45  \pm   0.01$  &$    -0.32  \pm  -0.02$  &B, M        \\
   6 &  J123923.66-113819.8 &  1.5 &$     0.37  \pm   0.12$  &$     0.02  \pm   0.05$  &$     0.12  \pm   0.06$  &$     0.16  \pm   0.07$  &$     0.06  \pm   0.06$  &$    -0.68  \pm  -0.02$  &$     0.05  \pm  -0.03$  &$    -0.48  \pm  -0.04$  &M           \\
   7 &  J123924.19-113838.0 &  0.9 &$     0.67  \pm   0.14$  &$     0.07  \pm   0.08$  &$     0.00  \pm   0.03$  &$     0.31  \pm   0.08$  &$     0.29  \pm   0.08$  &$    -0.14  \pm  -0.01$  &$     0.61  \pm   0.03$  &$    -0.04  \pm  -0.02$  &B, M, H2    \\
   8 &  J123925.15-113836.9 &  1.1 &$     0.40  \pm   0.11$  &$    -0.04  \pm   0.00$  &$     0.00  \pm   0.03$  &$     0.35  \pm   0.09$  &$     0.09  \pm   0.06$  &$    -0.55  \pm  -0.02$  &$     0.97  \pm   0.02$  &$    -0.58  \pm  -0.02$  &B, M        \\
   9 &  J123925.23-113408.8 &  1.4 &$     0.67  \pm   0.19$  &$     0.10  \pm   0.11$  &$     0.14  \pm   0.08$  &$     0.44  \pm   0.11$  &$    -0.01  \pm   0.07$  &$    -0.97  \pm  -0.01$  &$     0.30  \pm   0.02$  &$    -0.96  \pm  -0.02$  &B, M        \\
  10 &  J123926.66-113250.0 &  1.9 &$     0.44  \pm   0.20$  &$  0.24  \pm   0.15$  &$     0.07  \pm   0.06$  &$     0.27  \pm  0.09$  &$    -0.15  \pm   0.06$  &$    -0.97  \pm  -0.01$  &$-0.06  \pm  -0.02$  &$    -0.94  \pm  -0.03$  &M           \\
\enddata
\label{tab:sou}
\tablecomments{The definition of the bands:
0.4--0.7 (S1), 0.7--1 (S2), 1--2 (H1), and 2--6 (H2) keV. 
In addition, M=S2+H1 and F=S1+M+H2.
 Column (1): Generic source number. (2): 
{\sl Chandra} X-ray Observatory (unregistered) source name, following the
{\sl Chandra} naming convention and the IAU Recommendation for Nomenclature
(e.g., http://cdsweb.u-strasbg.fr/iau-spec.html). (3): Position 
uncertainty (1$\sigma$) calculated from the maximum likelihood centroiding.  (4): On-axis source F-band count rate --- the sum of the 
exposure-corrected count rates in the four
bands, based on the merged data of the three observations. In units of
 cts${\rm~s^{-1}}$. (5-8): Count rates in individual bands --- 0.4--0.7 (CR1), 0.7--1 (CR2), 1--2 (CR3), and 2--6 (CR4) keV. (9-11)
The hardness ratios defined as 
${\rm HR}=({\rm H2-M-S1})/({\rm H2+M+S1})$, ${\rm HR1}=({\rm M-S1})/{\rm M+S1}$ 
and ${\rm HR2}=({\rm H2-H1})/{\rm H2+H1}$ 
(12): The label ``B'', ``S1'', ``M'' or ``H2'' mark the band in 
which a source is detected. 
The full content of this table is available in the online journal.
}
  \end{deluxetable}

\begin{deluxetable}{cccccccc}
  \tabletypesize{\scriptsize}
  \tablecaption{List of sources detected from ObsID 1586\label{tab:sou_1586}}
  \tablewidth{0pt}
  \tablehead{
  \colhead{Source} &
  \colhead{CXOU Name} &
  \colhead{CR } &
  \colhead{CR1} &
  \colhead{CR2} &
  \colhead{CR3} &
  \colhead{CR4} &
  \colhead{Flag} \\
  \noalign{\smallskip}
  \colhead{(1)} &
  \colhead{(2)} &
  \colhead{(3)} &
  \colhead{(4)} &
  \colhead{(5)} &
  \colhead{(6)} &
  \colhead{(7)} &
  \colhead{(8)} 
  }
  \startdata
   1 &  J123919.60-113216.3 &$     1.99  \pm   0.53$  &$    -0.10  \pm   0.00$  &$     0.23  \pm   0.21$  &$     1.19  \pm   0.38$  &$     0.67  \pm   0.30$  &B, M        \\
   2 &  J123920.36-113258.0 &$     0.96  \pm   0.37$  &$    -0.08  \pm   0.00$  &$     0.07  \pm   0.13$  &$     0.88  \pm   0.31$  &$     0.08  \pm   0.16$  &M           \\
   3 &  J123920.73-113408.3 &$     0.97  \pm   0.37$  &$    -0.05  \pm   0.00$  &$     0.06  \pm   0.13$  &$     0.76  \pm   0.29$  &$     0.20  \pm   0.18$  &M           \\
   4 &  J123923.65-113816.4 &$     1.14  \pm   0.40$  &$    -0.04  \pm   0.00$  &$     0.56  \pm   0.29$  &$     0.41  \pm   0.22$  &$     0.21  \pm   0.16$  &B, M        \\
   5 &  J123924.31-113838.9 &$     0.79  \pm   0.31$  &$    -0.03  \pm   0.00$  &$    -0.01  \pm   0.00$  &$     0.43  \pm   0.23$  &$     0.41  \pm   0.21$  &B           \\
   6 &  J123925.24-113408.3 &$     1.43  \pm   0.46$  &$     0.22  \pm   0.26$  &$     0.24  \pm   0.19$  &$     0.90  \pm   0.31$  &$     0.07  \pm   0.12$  &B, M        \\
   7 &  J123926.63-113250.1 &$     0.68  \pm   0.33$  &$    -0.06  \pm   0.00$  &$     0.27  \pm   0.20$  &$     0.50  \pm   0.24$  &$    -0.02  \pm   0.09$  &M           \\
   8 &  J123928.68-113451.2 &$     1.79  \pm   0.54$  &$     0.51  \pm   0.38$  &$     0.39  \pm   0.24$  &$     0.79  \pm   0.29$  &$     0.10  \pm   0.12$  &M, B        \\
   9 &  J123929.43-114554.6 &$     6.68  \pm   1.13$  &$     1.20  \pm   0.63$  &$     0.54  \pm   0.33$  &$     2.47  \pm   0.63$  &$     2.48  \pm   0.61$  &H2, S1      \\
  10 &  J123929.78-114550.7 &$     5.94  \pm   0.95$  &$     0.95  \pm   0.50$  &$     0.43  \pm   0.27$  &$     3.02  \pm   0.63$  &$     1.55  \pm   0.44$  &B, M, S1    \\
\enddata
\tablecomments{The definition of the bands:
0.4--0.7 (S1), 0.7--1 (S2), 1--2 (H1), and 2--6 (H2) keV. 
In addition, M=S2+H1 and F=S1+M+H2.
 Column (1): Generic source number. (2): 
{\sl Chandra} X-ray Observatory (unregistered) source name, following the
{\sl Chandra} naming convention and the IAU Recommendation for Nomenclature
(e.g., http://cdsweb.u-strasbg.fr/iau-spec.html). (3): On-axis source F-band count rate --- the sum of the 
exposure-corrected count rates in the four
bands, based on the merged data of the three observations. In units of
 cts${\rm~s^{-1}}$. (4-7): Count rates in individual bands --- 0.4--0.7 (CR1), 0.7--1 (CR2), 1--2 (CR3), and 2--6 (CR4) keV. 
(8): The label ``B'', ``S1'', ``M'' or ``H2'' mark the band in 
which a source is detected. 
The full content of this table is available in the online journal.
}
  \end{deluxetable}

\begin{deluxetable}{cccccccc}
  \tabletypesize{\scriptsize}
  \tablecaption{List of sources detected from ObsID 9532\label{tab:sou_9532}}
  \tablewidth{0pt}
  \tablehead{
  \colhead{Source} &
  \colhead{CXOU Name} &
  \colhead{CR } &
  \colhead{CR1} &
  \colhead{CR2} &
  \colhead{CR3} &
  \colhead{CR4} &
  \colhead{Flag} \\
  \noalign{\smallskip}
  \colhead{(1)} &
  \colhead{(2)} &
  \colhead{(3)} &
  \colhead{(4)} &
  \colhead{(5)} &
  \colhead{(6)} &
  \colhead{(7)} &
  \colhead{(8)} 
  }
  \startdata
   1 &  J123915.22-114048.0 &$     0.52  \pm   0.17$  &$     0.04  \pm   0.09$  &$     0.04  \pm   0.05$  &$     0.34  \pm   0.11$  &$     0.11  \pm   0.09$  &M           \\
   2 &  J123918.61-113345.1 &$     1.14  \pm   0.28$  &$     0.64  \pm   0.24$  &$     0.05  \pm   0.06$  &$     0.30  \pm   0.10$  &$     0.14  \pm   0.09$  &B, S1       \\
   3 &  J123919.64-113214.7 &$     2.24  \pm   0.29$  &$     0.08  \pm   0.13$  &$     0.45  \pm   0.14$  &$     0.96  \pm   0.16$  &$     0.75  \pm   0.15$  &B, M, H2    \\
   4 &  J123920.13-113629.0 &$     0.83  \pm   0.21$  &$     0.24  \pm   0.15$  &$     0.13  \pm   0.07$  &$     0.33  \pm   0.10$  &$     0.14  \pm   0.08$  &M, B        \\
   5 &  J123920.62-113915.7 &$     0.83  \pm   0.17$  &$     0.05  \pm   0.08$  &$     0.05  \pm   0.05$  &$     0.55  \pm   0.12$  &$     0.18  \pm   0.08$  &B, M        \\
   6 &  J123920.64-113259.4 &$     0.41  \pm   0.14$  &$    -0.07  \pm   0.00$  &$     0.03  \pm   0.05$  &$     0.32  \pm   0.10$  &$     0.13  \pm   0.08$  &M           \\
   7 &  J123920.66-113407.7 &$     1.00  \pm   0.20$  &$     0.03  \pm   0.08$  &$     0.24  \pm   0.10$  &$     0.44  \pm   0.11$  &$     0.28  \pm   0.10$  &B, M        \\
   8 &  J123924.16-113837.8 &$     0.77  \pm   0.16$  &$     0.11  \pm   0.09$  &$     0.02  \pm   0.03$  &$     0.30  \pm   0.09$  &$     0.34  \pm   0.09$  &B, M, H2    \\
   9 &  J123925.17-113408.7 &$     0.83  \pm   0.21$  &$     0.14  \pm   0.12$  &$     0.13  \pm   0.08$  &$     0.38  \pm   0.11$  &$     0.18  \pm   0.09$  &M, B        \\
  10 &  J123925.19-113836.5 &$     0.46  \pm   0.12$  &$    -0.02  \pm   0.00$  &$     0.02  \pm   0.03$  &$     0.33  \pm   0.09$  &$     0.14  \pm   0.06$  &B, M        \\
\enddata
\tablecomments{The definition of the bands:
0.4--0.7 (S1), 0.7--1 (S2), 1--2 (H1), and 2--6 (H2) keV. 
In addition, M=S2+H1 and F=S1+M+H2.
 Column (1): Generic source number. (2): 
{\sl Chandra} X-ray Observatory (unregistered) source name, following the
{\sl Chandra} naming convention and the IAU Recommendation for Nomenclature
(e.g., http://cdsweb.u-strasbg.fr/iau-spec.html). (3): On-axis source F-band count rate --- the sum of the 
exposure-corrected count rates in the four
bands, based on the merged data of the three observations. In units of
 cts${\rm~s^{-1}}$. (4-7): Count rates in individual bands --- 0.4--0.7 (CR1), 0.7--1 (CR2), 1--2 (CR3), and 2--6 (CR4) keV. 
(8): The label ``B'', ``S1'', ``M'' or ``H2'' mark the band in 
which a source is detected. 
The full content of this table is available in the online journal.
}
  \end{deluxetable}

\begin{deluxetable}{cccccccc}
  \tabletypesize{\scriptsize}
  \tablecaption{List of sources detected from ObsID 9533\label{tab:sou_9533}}
  \tablewidth{0pt}
  \tablehead{
  \colhead{Source} &
  \colhead{CXOU Name} &
  \colhead{CR } &
  \colhead{CR1} &
  \colhead{CR2} &
  \colhead{CR3} &
  \colhead{CR4} &
  \colhead{Flag} \\
  \noalign{\smallskip}
  \colhead{(1)} &
  \colhead{(2)} &
  \colhead{(3)} &
  \colhead{(4)} &
  \colhead{(5)} &
  \colhead{(6)} &
  \colhead{(7)} &
  \colhead{(8)} 
  }
  \startdata
   1 &  J123937.25-113846.5 &$     0.32  \pm   0.10$  &$    -0.03  \pm   0.00$  &$     0.07  \pm   0.06$  &$     0.01  \pm   0.02$  &$     0.27  \pm   0.08$  &B, H2       \\
   2 &  J123937.74-114031.2 &$     2.41  \pm   0.28$  &$     0.14  \pm   0.14$  &$     0.44  \pm   0.13$  &$     0.98  \pm   0.16$  &$     0.85  \pm   0.13$  &B, M, H2    \\
   3 &  J123938.79-113851.6 &$     1.23  \pm   0.23$  &$     0.16  \pm   0.13$  &$     0.68  \pm   0.16$  &$     0.33  \pm   0.09$  &$     0.07  \pm   0.05$  &B, M        \\
   4 &  J123938.92-113728.0 &$     2.14  \pm   0.29$  &$     0.30  \pm   0.18$  &$     0.17  \pm   0.09$  &$     1.00  \pm   0.17$  &$     0.67  \pm   0.12$  &B, M, H2    \\
   5 &  J123939.38-113810.0 &$     0.85  \pm   0.16$  &$     0.07  \pm   0.09$  &$     0.14  \pm   0.07$  &$     0.46  \pm   0.10$  &$     0.18  \pm   0.06$  &B, M, H2    \\
   6 &  J123942.40-113656.1 &$     1.19  \pm   0.17$  &$    -0.01  \pm   0.00$  &$     0.19  \pm   0.08$  &$     0.80  \pm   0.13$  &$     0.21  \pm   0.06$  &B, M, H2    \\
   7 &  J123943.20-113644.8 &$     0.27  \pm   0.08$  &$    -0.01  \pm   0.00$  &$     0.06  \pm   0.05$  &$     0.10  \pm   0.05$  &$     0.11  \pm   0.05$  &B, M        \\
   8 &  J123943.62-114033.6 &$     3.72  \pm   0.36$  &$     0.53  \pm   0.22$  &$     0.55  \pm   0.14$  &$     1.52  \pm   0.19$  &$     1.13  \pm   0.15$  &B, M, H2, S \\
   9 &  J123943.69-113502.9 &$     0.78  \pm   0.16$  &$     0.15  \pm   0.11$  &$     0.06  \pm   0.05$  &$     0.43  \pm   0.10$  &$     0.15  \pm   0.05$  &B, M, H2    \\
  10 &  J123944.17-113600.6 &$     3.18  \pm   0.30$  &$     0.29  \pm   0.15$  &$     0.58  \pm   0.14$  &$     1.56  \pm   0.18$  &$     0.76  \pm   0.11$  &B, M, H2, S \\
\enddata
\tablecomments{The definition of the bands:
0.4--0.7 (S1), 0.7--1 (S2), 1--2 (H1), and 2--6 (H2) keV. 
In addition, M=S2+H1 and F=S1+M+H2.
 Column (1): Generic source number. (2): 
{\sl Chandra} X-ray Observatory (unregistered) source name, following the
{\sl Chandra} naming convention and the IAU Recommendation for Nomenclature
(e.g., http://cdsweb.u-strasbg.fr/iau-spec.html). (3): On-axis source F-band count rate --- the sum of the 
exposure-corrected count rates in the four
bands, based on the merged data of the three observations. In units of
 cts${\rm~s^{-1}}$. (4-7): Count rates in individual bands --- 0.4--0.7 (CR1), 0.7--1 (CR2), 1--2 (CR3), and 2--6 (CR4) keV. 
(8): The label ``B'', ``S1'', ``M'' or ``H2'' mark the band in 
which a source is detected. 
The full content of this table is available in the online journal.
}
  \end{deluxetable}

\begin{deluxetable}{ccccccccccc}
\scriptsize
  \tabletypesize{\scriptsize}
  \tablecaption{List of Sombrero GC-LMXBs}
  \tablewidth{0pt}
  \tablehead{
  \colhead{Source} &
  \colhead{RA} &
  \colhead{DEC} &
  \colhead{m$_B$} &
  \colhead{m$_V$} &
  \colhead{m$_R$} &
  \colhead{f$_X$} \\
  \noalign{\smallskip}
  \colhead{(1)} &
  \colhead{(2)} &
  \colhead{(3)} &
  \colhead{(4)} &
  \colhead{(5)} &
  \colhead{(6)} &
  \colhead{(7)} \\
  }
 \startdata
  46 &  189.83178 &  -11.53749 &    22.57 &    22.57 &    21.24 &    1.765 \\
  51 &  189.83359 &  -11.60794 &    20.25 &    20.25 &    18.71 &    0.595 \\
  52 &  189.83528 &  -11.54961 &    20.78 &    20.78 &    19.16 &    0.313 \\
  68 &  189.83583 &  -11.65434 &    19.75 &    19.75 &    18.49 &    0.583 \\
  77 &  189.83624 &  -11.56892 &    21.42 &    21.42 &    20.21 &    0.812 \\
  80 &  189.84860 &  -11.63886 &    20.96 &    20.96 &    19.37 &    0.367 \\
  88 &  189.85081 &  -11.64391 &    22.70 &    22.70 &    21.29 &    0.668 \\
  98 &  189.85481 &  -11.64360 &    21.47 &    21.47 &    20.15 &    0.405 \\
 106 &  189.85516 &  -11.56913 &    21.91 &    21.91 &    20.28 &    0.669 \\
 113 &  189.86110 &  -11.54725 &    21.48 &    21.48 &    19.87 &    0.437 \\
 118 &  189.86230 &  -11.66899 &    21.94 &    21.94 &    20.32 &    6.937 \\
 120 &  189.86257 &  -11.67380 &    22.28 &    22.28 &    21.03 &    0.890 \\
 121 &  189.86925 &  -11.58109 &    20.45 &    20.45 &    18.86 &    1.553 \\
 133 &  189.87201 &  -11.64634 &    21.68 &    21.68 &    20.26 &    0.911 \\
 137 &  189.87323 &  -11.76461 &    22.41 &    22.41 &    20.86 &    2.691 \\
 138 &  189.87814 &  -11.58502 &    22.80 &    22.80 &    21.58 &    0.674 \\
 166 &  189.87899 &  -11.76787 &    21.78 &    21.78 &    20.11 &    1.003 \\
 176 &  189.87911 &  -11.58132 &    22.47 &    22.47 &    20.88 &    0.837 \\
 179 &  189.88255 &  -11.54371 &    22.92 &    22.92 &    21.42 &    0.573 \\
 181 &  189.88405 &  -11.53095 &    22.48 &    22.48 &    21.34 &    0.583 \\
 185 &  189.88541 &  -11.67121 &    21.25 &    21.25 &    19.74 &    0.848 \\
 190 &  189.88553 &  -11.66165 &    22.02 &    22.02 &    20.49 &    0.761 \\
 202 &  189.89408 &  -11.57478 &    21.98 &    21.98 &    20.57 &    0.548 \\
 208 &  189.89634 &  -11.66120 &    21.62 &    21.62 &    20.48 &    0.205 \\
 214 &  189.89830 &  -11.64166 &    19.59 &    19.59 &    18.32 &    0.625 \\
 219 &  189.90062 &  -11.67650 &    22.14 &    22.14 &    20.60 &    0.427 \\
 231 &  189.90477 &  -11.64626 &    21.30 &    21.30 &    20.06 &    0.274 \\
 243 &  189.90722 &  -11.67529 &    21.16 &    21.16 &    19.61 &    2.373 \\
 252 &  189.91069 &  -11.70504 &    22.13 &    22.13 &    20.65 &    0.382 \\
 263 &  189.91158 &  -11.64759 &    20.41 &    20.41 &    18.89 &    1.073 \\
 283 &  189.91218 &  -11.62436 &    22.36 &    22.36 &    20.90 &    2.294 \\
 288 &  189.91310 &  -11.79109 &    21.39 &    21.39 &    20.20 &    2.006 \\
 293 &  189.91411 &  -11.63610 &    21.67 &    21.67 &    20.43 &    0.583 \\
 297 &  189.91848 &  -11.57028 &    21.29 &    21.29 &    20.03 &    0.400 \\
 299 &  189.91915 &  -11.54925 &    21.39 &    21.39 &    19.77 &    2.608 \\
 305 &  189.91946 &  -11.52358 &    22.12 &    22.12 &    20.49 &    1.796 \\
 312 &  189.92301 &  -11.56445 &    23.00 &    23.00 &    21.47 &    0.224 \\
 326 &  189.92331 &  -11.58600 &    19.79 &    19.79 &    18.48 &    0.256 \\
 329 &  189.92456 &  -11.62370 &    21.24 &    21.24 &    19.75 &    0.132 \\
 340 &  189.92675 &  -11.61560 &    22.09 &    22.09 &    20.63 &    1.197 \\
 346 &  189.93008 &  -11.61235 &    22.28 &    22.28 &    20.65 &    0.381 \\
\enddata
\label{tab:GCLMXB}
\tablecomments{Column (1): Generic X-ray source number as in Table~\ref{tab:sou}. (2-3): celetial coordinates. (4-6) $B$, $V$, $R$ apparent magnitudes. (7) 0.3-8 keV intrinsic flux in units of 10$^{-14}{\rm~ergs~s^{-1}}$. 
 }
  \end{deluxetable}


\begin{references}
\reference{} Bajaja E., Dettmar R.-J., Hummel E., Wielebinski R. 1988,
A\&A, 202, 35
\reference{} Bell E.F., \& de Jong R.S. 2001, ApJ, 550, 212 
\reference{} Bendo G.J., et al. 2006, ApJ, 645, 134
\reference{} Brassington N.J., et al. 2008, ApJS, 179, 142
\reference{} Brassington N.J., et al. 2008, ApJS, 181, 605
\reference{} Bridges T.J., Rhode K.L., Zepf S.E., Freeman K.C. 2007,
ApJ, 658, 980
\reference{} Cash W. 1979, ApJ, 228, 939
\reference{} Clark G.W. 1975, ApJ, 199, L143
\reference{} Cutri R.M., et al. 2003, The IRSA 2MASS All-Sky Catalog of Point Sources, NASA/IPAC Infrared Science Archive, http://irsa.ipac.caltech.edu/applications/Gator
\reference{} Dickey J.M., Lockman F.J. 1990, \araa, 28, 215
\reference{} Di Stefano R., Kong A.K.H. 2004, ApJ, 609, 710
\reference{} Di Stefano R., Kong A.K.H., Van Dalfsen M.L., Harris W.E., Murray S. S., Delain K.M., 2003, ApJ, 599, 1067
\reference{} Ebrero J., Mateos S., Stewart G.C., Carrera F.J., Watson M.G., 2009, A\&A, 500, 749
\reference{} Gilfanov M. 2004, MNRAS, 349, 146
\reference{} Grimm H.-J., Gilfanov M., Sunyaev R. 2003, MNRAS, 339, 793
\reference{} Harris W.E, Spitler L.R., Forbes D.A., Bailin J. 2009,
MNRAS, submitted
\reference{} Hau G.K.T. et al. 2009, MNRAS, 394, L97
\reference{} Hickox R.C. \& Markevitch M. 2007, ApJ, 661, L117
\reference{} Hickox R.C. et al. 2009, ApJ, 696, 891
\reference{} Jarrett T.H., Chester T., Cutri R., Schneider S.E., Huchra J.P., 2003, \aj, 125, 525
\reference{} Kim E., et al. 2006, ApJ, 647, 276
\reference{} Kong A.K.H., Garcia M.R., Primini F.A., Murray S.S., Di Stefano, R., McClintock J.E. 2002, ApJ, 577, 738
\reference{} Kraft R.P., Kregenow J.M., Forman W.R., Jones C., Murray S.S. 2001, ApJ, 560, 675 
\reference{} Kulkarni S.R., Hut P., McMillan S. 1993, 364, 421
\reference{} Kundu A., Maccarone T.J., Zepf S. 2007, ApJ, 662, 525
\reference{} Lahav O., \& Saslaw W.C. 1992, ApJ, 396, 430
\reference{} Li Z., Wang Q.D, Hameed S. 2007 MNRAS, 376, 960
\reference{} Maccarone T.J., Kundu A., Zepf S., Rhode K.L. 2007, Nature, 445, 183
\reference{} Miyaji T. \& Griffiths R. 2002, ApJ, 564, L5
\reference{} Monet D., et al. 2003, AJ, 125, 984
\reference{} Moretti A., Campana S., Lazzati D., Tagliaferri G., 2003, ApJ, 588, 696
\reference{} Paczy$\acute{n}$ski B. 1990, ApJ, 348, 485
\reference{} Park T., Kashyap V.L., Siemiginowska A., van Dyk D.A., Zezas A., Heinke C., Wargelin B.J. 2006, ApJ, 652, 610
\reference{} Peebles P.J.E. 1980, The Large-Scale Structure of the Universe (Princeton: Princeton Univ. Press)
\reference{} Pooley D., \& Hut P. 2006, ApJ, 646, L143   
\reference{} Revnivtsev M., Churazov E., Sazonov S., Forman W., Jones
C. 2007, A\&A, 490, 37
\reference{} Revnivtsev M., Churazov E., Sazonov S., Forman W., Jones
C. 2008, A\&A, 473, 783 
\reference{} Revnivtsev M., Sazonov S., Gilfanov M., Churazov E.,
Sunyaev R. 2006, A\&A, 452, 169
\reference{} Rhode K., \& Zepf S.E. 2004, AJ, 127, 302
\reference{} Posson-Brown, J., Raychaudhury S., Forman W., Donnelly,
R. H., Jones C. 2009, ApJ, 695, 1094
\reference{} Sarazin C.L., Irwin J.A., Bregman J.N. 2000, ApJ, 544, L101
\reference{} Sigurdsson S., \& Hernquist L. 1993, Nature, 364, 423
\reference{} Sivakoff G. et al. 2008, astro-ph/0806.0626
\reference{} Spitler L.R., Larsen S.S., Strader J., Brodie J.P., Forbes D.A., Beasley M.A. 2006, ApJ, 132, 1593
\reference{} Spitler L.R., Forbes D.A., Beasley M.A. 2008, MNRAS, 389, 1150 
\reference{} Tennant A.F., Wu K., Ghosh K.K., Kolodziejczak J.J., Swartz D.A. 2001, ApJ, 549, L43
\reference{} Vikhlinin A., \& Forman W. 1995, ApJ, 451, 553
\reference{} Voss R., \& Gilfanov M. 2007, A\&A, 468, 49
\reference{} Wang Q.D., 2004, ApJ, 612, 159
\reference{} Zepf S.E., et al. 2008, ApJ, 683, L139
\reference{} Zezas A., Hernquist L., Fabbiano G., Miller J. 2003, ApJ, 599, L73
\reference{} Zuo Z.-Y., Li X.-D., Liu X.-W. 2008, MNRAS, 387, 121
\end{references}
\end{document}